\documentclass[prb,showpacs,twocolumn,superscriptaddress,floatfix,amsmath]{revtex4}
 \usepackage[latin1]{inputenc}
 \usepackage[english]{babel}
 \usepackage{graphicx}
 \usepackage{psfrag}
 \usepackage{amssymb}
 \usepackage{amsmath}
 \usepackage{amscd}
 \usepackage{eucal}
 \usepackage{color}
 \usepackage{bm}
 
 \newcommand{\ignore}[1]{\relax}

 \usepackage{epstopdf}
 \DeclareMathAlphabet{\mathpzc}{OT1}{pzc}{m}{it}

 \newcommand{\pll}{\parallel}
 
 \newcommand{\e}{{\rm e}}
 \newcommand{\rmd}{{\rm d}}
 \newcommand{\rmi}{{\rm i}}

 \newcommand{\half}{{\textstyle{\frac{1}{2}}}}

 \newcommand{\al}{\alpha}
 
 \newcommand{\de}{\delta}
 
 \newcommand{\eps}{\epsilon}
 
 \newcommand{\om}{\omega}

 \newcommand{\tEo}{\tau_{\rm E}^{\rm op}}
 \newcommand{\tEc}{\tau_{\rm E}^{\rm cl}}
 \newcommand{\tE}{\tau_{\rm E}}

 \newcommand{\tD}{\tau_{\rm D}}

 \newcommand{\doublelangle}{\langle\!\langle}
 \newcommand{\doublerangle}{\rangle\!\rangle}

 \newcommand{\bdoublelangle}{\big\langle\! \big\langle}
 \newcommand{\bdoublerangle}{\big\rangle\! \big\rangle}

 \newcommand{\Bdoublelangle}{\Big\langle\!\! \Big\langle}
 \newcommand{\Bdoublerangle}{\Big\rangle\!\! \Big\rangle}

 \newcommand{\Bggdoublelangle}{\Bigg\langle\!\!\! \Bigg\langle}
 \newcommand{\Bggdoublerangle}{\Bigg\rangle\!\!\! \Bigg\rangle}

 \definecolor{DarkGreen}{rgb}{0,0.7,0}

\begin{document}

 \title{Dephasing in quantum chaotic transport: a semiclassical approach}

 \author{Robert S. Whitney}
 \affiliation{Institut Laue-Langevin,
 6 rue Jules Horowitz, BP 156, 38042 Grenoble, France}

 \author{Philippe Jacquod}
 \affiliation{Physics Department, University of Arizona, 
 1118 E. 4$^{\rm th}$ Street, Tucson, AZ 85721, USA}

 \author{Cyril Petitjean}
 \affiliation{D\'epartement de Physique Th\'eorique,
 Universit\'e de Gen\`eve, CH-1211 Gen\`eve 4, Switzerland}
 \affiliation{Institut I -- Theoretische Physik,
 Universit\"at Regensburg, Universit\"atsstrasse 31, 
 D-93040 Regensburg, Germany}

 \date{October 26, 2007}
 \begin{abstract}
 We investigate the effect of dephasing/decoherence on
 quantum transport through open chaotic ballistic conductors in the
 semiclassical limit of small Fermi wavelength to system size ratio,
 $\lambda_{\rm F}/L \ll 1$. We use the trajectory-based semiclassical theory
 to study a two-terminal chaotic dot
 with decoherence originating from:
 (i)  an external closed quantum chaotic environment, 
 (ii) a classical source of noise,
 (iii) a voltage probe, i.e. an additional current-conserving terminal.
 We focus on the pure dephasing regime, where the coupling to the 
 external source of dephasing is so weak that it does not induce
 energy relaxation. In addition to the universal algebraic suppression 
 of weak localization, we find an exponential suppression of weak-localization
 $\propto \exp[-\tilde{\tau}/\tau_\phi]$, 
 with the dephasing rate $\tau_\phi^{-1}$.
 The parameter $\tilde{\tau}$ depends strongly on the source of dephasing.
 For a voltage probe, $\tilde{\tau}$ is of order the Ehrenfest time $\propto \ln [L/\lambda_{\rm F}]$.
 In contrast, for a chaotic environment or a classical source of noise,
 it has the correlation length $\xi$ of the 
 coupling/noise potential replacing the Fermi wavelength $\lambda_{\rm F}$. 
 We explicitly show that the Fano factor for shot noise is unaffected by decoherence.
 We connect these results to earlier works on
 dephasing due to electron-electron interactions, and numerically confirm our
 findings.
 \end{abstract}
 \pacs{05.45.Mt,74.40.+k,73.23.-b,03.65.Yz}
 % 05.45.Mt Quantum chaos; semiclassical methods
 % 74.40.+k Fluctuations (noise, chaos, nonequilibrium superconductivity,
 % localization, etc.)
 % 73.23.-b Electronic transport in mesoscopic systems
 % 03.65.Yz Decoherence; open systems; quantum statistical methods
 \maketitle 

  %~~~~~~~~~~~~~~~~~~~~~~~~~~~~~~~~~~~~~~~~~~~~~~~~~~~~~~~~~~~~~~%

 %~~~~~~~~~~~~~~~~~~~~~~~~~~~~~~~INTRODUCTION~~~~~~~~~~~~~~~~~~~~~~~~~~~~~%
 \section{INTRODUCTION }

 \subsection{Dephasing in the universal regime}

 Electronic systems in the mesoscopic regime are ideal testing-grounds for 
 investigating the quantum-to-classical 
 transition from a microscopic coherent world
 (where quantum interference effects prevail) 
 to a macroscopic classical world~\cite{Joo03}. 
 On one hand, their size is intermediate between macroscopic 
 and microscopic (atomic) systems, on the other hand, today's experimental
 control over their design and precision of measurement allows one 
 to investigate
 them in regimes ranging from almost fully coherent to purely 
 classical~\cite{Sto92,Akk07,Imr02}. The extent to which quantum
 coherence is preserved in these systems is usually determined by the ratio
 $\tau_\phi/\tau_{\rm cl}$ of the dephasing time $\tau_\phi$ to some
 relevant classical time scale $\tau_{\rm cl}$. For instance, $\tau_{\rm cl}$
 can be the traversal time through one arm of a two-path
 interferometer~\cite{Ste90,See03,Jac05}, or
 the average dwell time spent inside 
 a quantum dot~\cite{Alt82,Cha86,Bar95,Brou97,Ale99}. In a given experimental 
 set-up, $\tau_\phi$ can often be tuned
 from $\tau_\phi > \tau_{\rm cl}$ (quantum coherent regime) 
 to $\tau_\phi \ll \tau_{\rm cl}$ (purely classical regime)
 by varying externally applied voltages or the temperature of the
 sample.

 Coherent effects abound in mesoscopic physics, the most important of them
 being weak-localization, universal conductance fluctuations and 
 Aharonov-Bohm interferences in transport,
 as well as persistent currents~\cite{Akk07,Imr02,Sto92}. The disappearance
 of these effects as dephasing processes are turned on has raised lots of  
 theoretical~\cite{Alt82,Cha86,Bar95,Brou97,Ale99,
 Fukuyama83,Ale96,Narozhny02,
 But86,Jon96,Vav99,See01, Eck02,Pol03,Tex05,
 Two04.2,Bro06-quasi-ucf, petitjean06} and 
 experimental~\cite{Hui98,Marcus-expt-microwaves,
 Heiblum,Yev00,Han01,Kob02,Pie03,vanderwiel} 
 interest. Focusing on transport through ballistic systems, 
 dephasing is usually investigated using mostly phenomenological models 
 of dephasing~\cite{But86,Jon96,Vav99,See01,Pol03,Tex05},
 the most successful of which are the voltage-probe and dephasing-lead
 models~\cite{But86,Jon96}. In these models, a cavity is connected
 to two external, L (left) and R (right) 
 transport leads, carrying $N_{\rm R}$ and $N_{\rm L}$ transport
 channels respectively. 
 Dephasing is modeled by connecting a third ``fictitious'' lead to
 the system, with a voltage set such that no 
 current flows through
 it on average. Electrons leaving the system through this third lead
 are thus re-injected at some later time, with a randomized phase
 (and randomized momentum). 
 These models of dephasing
 present the significant 
 advantage that the standard scattering approach to transport
 can be applied as in fully coherent systems, once it is properly extended to 
 account for the presence of the third lead. 

 Using random matrix theory (RMT), the voltage/dephasing
 probe models\cite{Hei07}
 predict an algebraic suppression of the 
 weak-localization contribution to the conductance (in units of $2e^2/h$)~\cite{Been97}, 
 \begin{eqnarray}
 g^{\rm wl}_{\rm RMT} &=& -\frac{N_{\rm R} N_{\rm L}}{[N_{\rm R}+N_{\rm L}]^2}
 \; [1+\tau_{\rm D}/\tau_\phi]^{-1}, 
 \label{eq:dephasing-wl-without-Ehrenfest}
 \end{eqnarray}
 where $-N_{\rm R} N_{\rm L}/[N_{\rm R}+N_{\rm L}]^2$ is the
 weak-localization correction in the absence of dephasing.
 Similarly, universal conductance fluctuations become~\cite{Brou97}
 \begin{eqnarray}
 \delta g^2_{\rm RMT} &=& \frac{2}{\beta} 
 \frac{N_{\rm R}^2 N^2_{\rm L}}{[N_{\rm R}+N_{\rm L}]^4} \;
 [1+\tau_{\rm D}/\tau_\phi]^{-2},
 \label{eq:dephasing-ucf-without-Ehrenfest}
 \end{eqnarray}
 and are thus damped below their value
 $(2/\beta) N^2_{\rm R} N^2_{\rm L}/[N_{\rm R}+N_{\rm L}]^2$ 
 (in units of $4e^4/h^2$) for fully coherent systems
 with ($\beta=1$) or without ($\beta=2$) time reversal symmetry.
 For a fictitious lead connected to a two-dimensional cavity
 (a lateral quantum dot) via a point contact of 
 transparency $\rho$ and carrying $N_3$ channels one has
 $\tau_\phi = m A/\hbar \rho N_3$,
 with the electron mass $m$ and the area $A$ of the cavity.
 Similarly, the dwell time through the cavity is given 
 by $\tau_{\rm D} = m A/\hbar (N_{\rm L}+N_{\rm R})$.

 The dephasing and voltage probe models 
 account for dephasing at the phenomenological level only, without
 reference to the microscopic processes leading to 
 dephasing. At sufficiently low temperature, it is accepted that the 
 dephasing arises dominantly from electronic interactions, which,
 in diffusive systems, can be well modeled  
 by a classical noise potential~\cite{Alt82, Cha86}. Remarkably enough,
 this approach reproduces the RMT results of 
 Eqs.~\eqref{eq:dephasing-wl-without-Ehrenfest} and 
 \eqref{eq:dephasing-ucf-without-Ehrenfest} with $\tau_\phi$
 set by the noise power. These results are moreover quite robust in diffusive
 systems. They are essentially insensitive to most noise-spectrum details, 
 and hold for various sources of noise such as electron-electron and
 electron-phonon interactions, or external microwave fields. For this
 reason it is often assumed that dephasing is system-independent,
 and exhibits a character of universality well described by the RMT
 of transport applied to the dephasing/voltage probe models~\cite{Been97}.

 \subsection{Departure from RMT universality} 

 According to the Bohigas-Giannoni-Schmit surmise~\cite{BGS},
 closed chaotic systems exhibit statistical
 properties of hermitian RMT~\cite{Meh91} in the short wavelength limit. 
 Opening up the system,
 transport properties derive from the corresponding scattering matrix, which
 is determined by both the Hamiltonian of the closed system and its coupling
 to external leads~\cite{RMT-transport}. It has been shown that for not 
 too strong coupling, and when the Hamiltonian matrix of the closed system
 belongs to one of the Gaussian
 ensembles of random hermitian matrices, the corresponding scattering matrix
 is an element of one of the circular ensembles of unitary random 
 matrices~\cite{Lew91}. One thus expects that, in the semiclassical
 limit of large ratio $L/\lambda_{\rm F}$ of the system size to 
 Fermi wavelength, transport properties of quantum chaotic ballistic systems 
 are well described by the RMT of 
 transport. This surmise has recently been verified 
 semiclassically~\cite{Heu06}.  

 The regime of validity of RMT is generally bounded by the existence of
 finite time scales, however, and
 it was noticed by Aleiner and Larkin that,
 while the dephasing time $\tau_{\phi}$ gives the long time cut-off for 
 quantum interferences, a new {\it Ehrenfest time} scale appears in
 quantum chaotic system in the deep semiclassical limit, which determines 
 the short-time onset
 of these interferences~\cite{Ale96}.
 The Ehrenfest time $\tE$ corresponds to the time it takes for the underlying
 chaotic classical dynamics to stretch an initially localized wavepacket
 to a macroscopic, classical length scale. In open cavities, the latter 
 can be either the system size $L$ or the width $W$
 of the opening to the leads. Accordingly, one can define
 the closed cavity, $\tEc= \lambda^{-1} \ln[ L/\lambda_{\rm F}]$, and the open 
 cavity  Ehrenfest time, 
 $\tEo= \lambda^{-1} \ln[ W^2/\lambda_{\rm F}L]$~\cite{Vav03,Scho05}.
 The emergence of a finite $\tE$ strongly affects quantum effects
 in transport, and recent analytical and numerical investigations 
 of quantum chaotic systems have shown that weak 
 localization~\cite{Ale96,Ada03,Bro06-quasi-wl,Jac06,Bro06-cbs}
 and shot noise~\cite{agam,Two03,wj2004,wj2005-fano}
 are exponentially suppressed $\propto \exp[-\tE/\tD]$ in the absence of
 dephasing ($\tau_\phi \rightarrow \infty$).   
 Interestingly enough, the deep semiclassical limit of finite $\tE$
 sees the emergence
 of a quantitatively dissimilar behavior of weak-localization and
 quantum parametric conductance fluctuations, the latter exhibiting no
 $\tE$-dependence in the absence of 
 dephasing~\cite{JacqSukho04,Two04,Bro06-ucf,caveatucf}.
 These results are not captured by RMT, instead one has to rely
 on quasiclassical approaches~\cite{Ale96,Bro06-quasi-wl} or semiclassical
 methods~\cite{Ada03,Jac06,Bro06-cbs,Ric02,Whi07} to derive 
 them.

 \subsection{Dephasing in the deep semiclassical limit} 

 The behavior of quantum corrections to transport at finite $\tE$ in
 the presence of dephasing was briefly investigated analytically 
 in Ref.~[\onlinecite{Ale96}], for a model of classical noise with 
 large angle scattering, and numerically in Ref.~[\onlinecite{Two04.2}],
 for the dephasing lead model with a tunnel barrier. 
 Intriguingly enough, the two approaches
 delivered the same result, that quantum effects are exponentially
 suppressed $\propto \exp[-\tE/\tau_\phi]$. This suggested that dephasing
 retains a character of universality even in the deep semiclassical limit.
 More recent investigations have however
 showed that at finite Ehrenfest time, 
 Eq.~(\ref{eq:dephasing-wl-without-Ehrenfest}) 
 becomes~\cite{petitjean06} (see also Ref.~[\onlinecite{Bro06-quasi-ucf}])
 \begin{eqnarray}
 g^{\rm wl} &=& 
  -\frac{N_{\rm R} N_{\rm L}}{[N_{\rm R}+N_{\rm L}]^2} \;
 \frac{\exp[-(\tEc/\tD + \tilde{\tau}/\tau_\phi)]}{1+\tau_{\rm D}/\tau_\phi},
 \label{eq:wl-with-Ehrenfest}
 \end{eqnarray}
 with a strongly system-dependent time scale $\tilde{\tau}$.
 Ref.~[\onlinecite{petitjean06}] showed that, for the dephasing lead model, 
 $\tilde{\tau}=\tEc + (1-\rho)\tEo$ in terms of
 the transparency $\rho$ of the contacts to the leads, which 
 provides theoretical understanding for the numerical findings of Ref.~[\onlinecite{Two04.2}]. 
 If however one considers a system-environment model, where
 the environment is mimicked by electrons in a nearby 
 closed quantum chaotic dot, one has
 $\tilde{\tau} = \tau_\xi$,
 where 
 \begin{eqnarray}
 \tau_\xi =\lambda^{-1}\ln [(L/\xi)^2],
 \label{eq:tau_xi}
 \end{eqnarray} 
 in terms of the
 correlation length $\xi$ of the inter-dot interaction potential. 

 On the experimental front, 
 an exponential suppression $\propto \exp[-T/T_c]$
 of weak-localization with temperature has
 been reported in Ref.~[\onlinecite{Yev00}]. Taking $\tau_\phi \sim T^{-1}$
 as for dephasing by electronic interactions in two-dimensional 
 diffusive systems, this result was interpreted as
 the first experimental confirmation of Eq.~(\ref{eq:wl-with-Ehrenfest}). There is
 no other theory for such an exponential behavior of weak-localization, however, the temperature
 range over which this experiment has been performed makes it
 unclear whether the ballistic~\cite{Fukuyama83,Narozhny02}, 
 $\tau_\phi \sim T^{-2}$,
 or the diffusive dephasing time determines the Ehrenfest time dependence 
 of dephasing (see the discussion in Ref.~[\onlinecite{Bro06-quasi-ucf}]).

 %%%%%%%%%%%%%%%%%%%%%%%%%%%%%%%%%%%%%%%%%%%%%%%%%%
 \begin{table*}
 \begin{tabular}{|l||c|c|c|} 
 \hline
 & 
 $\Big.$\ {\bf Weak localization}\ & \ {\bf Conductance fluctuations} & 
 \ {\bf Shot noise}\ 
 \\ 
 \hline \hline
 \ {\bf System with environment} 
 & $\Big. \tilde{\tau} = \tau_\xi $ 
 &  --- & 
 \\
 \cline{1-3}
 \ {\bf Classical noise} (microwave, etc)  & 
 $\Big. \tilde{\tau} = \tau_\xi$ & 
 $\tilde{\tau} \sim 0$,   Ref.~[\onlinecite{Bro06-quasi-ucf}] & no dephasing 
 \\
 \ {\bf e-e interactions within system} \ &  
 \ $\tilde{\tau} =\tau_E^{\rm cl} + \half \tau_\xi$ \ & 
 $\tilde{\tau} \sim \half \tau_E^{\rm cl}$, Ref.~[\onlinecite{Bro06-quasi-ucf}] &
 \\
 \ {\bf System-gate e-e interactions} \ & 
 $\Big. \tilde{\tau} = \tau_\xi$ &
 \ \ $\tilde{\tau} \sim \half \tau_\xi$, 
 follows from Ref.~[\onlinecite{Bro06-quasi-ucf}] \ &
 \\
 \hline
 \ {\bf Dephasing lead:} & & & 
 \\
 \ \qquad no tunnel barrier &
 $\tilde{\tau} = \tau_E^{\rm cl}$ &
 $\tilde{\tau}= 0$ & \ no dephasing, Ref.~[\onlinecite{vanLan97}]
 \\
 \ \qquad low transparency barrier &
 \ \ $\Big. \tilde{\tau}=\tau_E^{\rm op}+\tau_E^{\rm cl}\sim 2\tau_E^{\rm cl}$ \ &
  $\Big.  \tilde{\tau} \sim 2\tau_E^{\rm cl}$ 
 (numerics), Ref.~[\onlinecite{Two04.2}] &
 \\
 \hline
 \end{tabular}
 \caption{\label{table1}
 Summary of the known results to date 
 on the nature of the exponential term $\exp[-\tilde{\tau}/\tau_\phi]$
 in the dephasing [cf. Eq.~(\ref{eq:wl-with-Ehrenfest})].
 Results that are not referenced are obtained in the present article
 and in Ref.~[\onlinecite{petitjean06}].
 Here we list the value of $\tilde{\tau}$ for different transport
 quantities and different sources of dephasing, all in the pure dephasing
 regime (the phase-breaking regime of Ref.~[\onlinecite{deJon96}]). 
 The parameter $\xi$ differ slightly from system to
 system (see text for details) however it is always related to
 the correlation length of the interaction with the environment.
 The results of Ref.~[\onlinecite{Bro06-quasi-ucf}]
 neglect $\tau_\xi$-contributions (so ``0'' could indicate a
 $\tilde{\tau}$ of order $\tau_\xi$), they are also 
 only valid for $\tau_E^{\rm cl} \gg \tau_{\rm \phi}$.}
 \end{table*}
 %%%%%%%%%%%%%%%%%%%%%%%%%%%%%%%%%%%%%%%%%%%%%%%%%%%%%

 \subsection{Outline of this article}

 In the present article, we amplify on Ref.~[\onlinecite{petitjean06}] and  
 extend the analytical derivation of Eq.~(\ref{eq:wl-with-Ehrenfest}) 
 briefly presented there.
 We investigate three different models of 
 dephasing and show that the suppression of weak-localization corrections to 
 the conductance is strongly model-dependent. 
 First, we consider an external environment  modeled by a capacitively coupled, 
 closed quantum dot. We restrict ourselves  to the regime of pure dephasing, 
 where the environment does not alter the classical dynamics of the system. 
 Second, we discuss dephasing by a classical noise field. 
 Third, following Ref.~[\onlinecite{Whi07}], we provide a semiclassical treatment 
 of transport in the dephasing lead model. For these three models, we
 reproduce Eq.~(\ref{eq:wl-with-Ehrenfest}) and derive the exact
 dependence of $\tilde{\tau}$ on microscopic details of the models
 considered.
All our results are summarized in Table~\ref{table1}.

 The outline of this article goes as follows.  
 In Section ~\ref{SYS-ENV-SEC}, we present the treatment of the 
 system-environment model, focusing in particular on the construction of
 a new scattering approach to transport that incorporates the coupling
 to external degrees of freedom. We apply this formalism to a 
 model of an open quantum dot coupled to a second, closed quantum dot.
 We present a detailed calculation of 
 the Drude conductance and the weak-localization correction,
 including coherent-backscattering, which explicitly preserves the unitarity 
 of the $S$-matrix, and hence
 current conservation. This calculation is completed by a derivation 
 of the Fano factor, showing that,  in the pure dephasing limit, 
 shot noise is insensitive to dephasing to leading order.
 In Section~\ref{sect:classical}, we present a model of dephasing via 
 a classical noise field (such as microwave noise).
 We consider classical Johnson-Nyquist noise models
 of dephasing due to electron-electron interactions within the system, 
 and dephasing due to charge fluctuations on nearby gates.
 In Section ~\ref{DEPHAS-SEC} we present a trajectory based semiclassical
 calculation of conductance in the dephasing  lead model, both for
 fully transparent barriers and tunnel barriers. We also comment on
 dephasing in multiprobe configurations.
 Finally, Section ~\ref{NUM-SEC} is devoted to numerical simulations
 confirming our analytical results. Summary and conclusions are presented in 
 Section ~\ref{CONCL-SEC}, while technical details are presented in the Appendix.

  %~~~~~~~~~~~~~~TRANSPORT THEORY FOR A SYSTEM-ENVIRONMENT~~~~~~~~~~~~~~~%
 \section{TRANSPORT THEORY FOR A SYSTEM WITH ENVIRONMENT  }\label{SYS-ENV-SEC}

 In the scattering approach to transport, the system is assumed fully coherent
 and all dissipative processes occur in the leads~\cite{Butt92}.   
 Apart from its coupling to the leads, the system is isolated. Here we
 extend this formalism to include coupling to external degrees of freedom
 in the spirit of the standard theory of decoherence. The coupling to
 an environment can induce dephasing and relaxation. Here, we restrict
 ourselves to pure dephasing, where the system-environment coupling
 does not induce energy, nor momentum relaxation in the system. In
 semiclassical language, we assume that classical trajectories supporting
 the electron dynamics are not modified by this coupling. 

 The starting point of the standard theory of decoherence
 is the total density matrix $\eta_{\rm tot}$  that includes both system and 
 environmental degrees of freedom~\cite{Joo03}.  
 The observed properties of the  system alone are contained in 
 the reduced density matrix $\eta_{\rm sys} $, 
 obtained from $\eta_{\rm tot}$ by tracing over the environmental 
 degrees of freedom. This procedure is probability conserving, 
 ${\rm Tr} \left[\eta_{\rm sys} \right]=1 $, but it renders the time-evolution 
 of $\eta_{\rm sys}$ non-unitary, and in particular, the off-diagonal
 elements of $\eta_{\rm sys}$ decay with time. This can be quantified
 by the basis independent purity~\cite{Zur03},
 $0 \le {\rm Tr} \left[\eta_{\rm sys}^2\right] \le 1$,
 which remains equal to one only in the absence of environment.
 We generalize this standard approach to the transport problem.

 %--------------------------------------------------------------------------------------------%
 \subsection{The scattering formalism in the presence of an environment } 

 %%%%%%%%%%%%%%%%%%%%%%%%%%%%%%%%%%%%%%%%% %%%%%%%
 \begin{figure}
 \begin{center}
 \includegraphics[width=7cm]{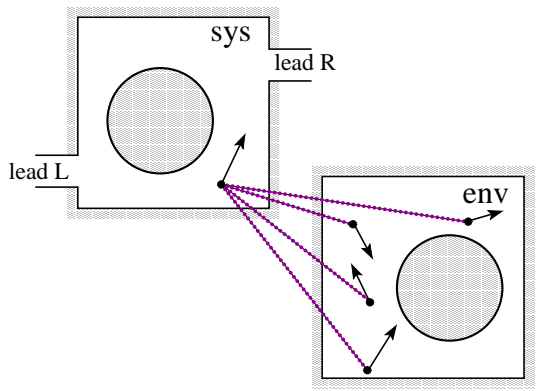} \hspace{0.3cm}
 \caption{\label{fig:env_model} 
 Schematic of the system-environment  model.
 The system is an open quantum dot that is coupled to an environment in the 
 shape of a second, closed dot.}
 \end{center}
 \end{figure}
 %%%%%%%%%%%%%%%%%%%%%%%%%%%%%%%%%%%%%%%%% %%%%%%%

 We consider two capacitively coupled chaotic cavities  
 as sketched in Fig.~\ref{fig:env_model}.       
 The first one is the system (sys), an open,  
 two-dimensional quantum dot, ideally connected to two external leads. 
 The second one is a closed quantum dot, which plays the role of  
 an environment (env).   
 The two dots are capacitively coupled, and in particular, they do not exchange 
 particles. Thus current through the system is conserved.
 We require that the size of the contacts between the open system and the leads is 
 much smaller than the perimeter of the system cavity  but is still 
 semiclassically large, so that the number of transport channels satisfies 
 $1 \ll N_{\rm L,\,R}  \ll L/\lambda_F$. This ensures that the chaotic dynamics 
 inside the dot has enough time to develop, $\lambda \tau_{\rm D} \gg 1$, with
 the classical Lyapunov exponent $\lambda$.  
 Electrons in the leads do not interact with the second dot.     
 Few-electron double-dot systems have recently 
 been the focus of intense experimental efforts~\cite{Wiel03}.
 Parallel geometries, of direct relevance to the present work, have been
 investigated in Refs.~[\onlinecite{Hof95,Adou99}].

 The total system is described by the following 
 Hamiltonian 
 \begin{equation}\label{2hamiltonian}
 {\cal H} = H_{\rm sys} + H_{\rm env} + {\cal U}.
 \end{equation}
 Inside each cavity the chaotic dynamics is generated by the corresponding 
 one-particle Hamiltonian $ H_{\rm sys,\,env}$.   
 We only specify that the capacitive coupling  potential ${\cal U}$  is a 
 smooth function of  the distance between the particles. It is characterized 
 by its magnitude $U$ and its correlation length $\xi$ such that its typical 
 gradient is $U/\xi$. Physically, $\xi$ is determined by the electrostatic
 environment of the system, such as electric charges on the gates defining the 
 dots and the amount of depletion of the electrostatic confinement 
 potential between the gates and the inversion layer
 in semiconductor heterostructures. Generally speaking, $U$ and $\xi$ are
 independent parameters and can have different values in different systems,
 and might even be tuned by applying external backgate voltages on a given system.

 In the standard scattering approach, the transport properties of the system 
 derive from its $(N_{\rm L} + N_{\rm R}) \times (N_{\rm L} + N_{\rm R})$ scattering matrix~\cite{But86}
 \begin{eqnarray}\label{blocks}
 {\hat {\bf  S}}= \left( \begin{array}{ll}
 {\bf s}_{\rm LL} & {\bf s}_{\rm RL} \\
 {\bf s}_{ \rm LR}  & {\bf s}_{\rm RR}
 \end{array}\right),
 \end{eqnarray}
 which we write in terms of 
 transmission (${\bf t} = {\bf s}_{\rm LR}$) and 
 reflection (${\bf r}={\bf s}_{\alpha,\alpha} $, $\alpha \in \{{\rm L,\,R}\}$) 
 matrices. From ${\hat {\bf S}}$, the system's dimensionless
 conductance (conductance in units of $2e^2/h$) is given by 
 \begin{eqnarray}\label{buttform}
 g={\rm Tr}({\bf t}^\dagger {\bf t}).
 \end{eqnarray}
 To include coupling to an environment in the scattering 
 approach, we need to define an extended scattering  matrix ${\mathbb S}$ that  
 includes the external degrees of freedom. This is formally done in 
 Appendix~\ref{extendS}, and our starting 
 point is Eq.~(\ref{conductance}) for the case of an initial
 product density matrix $\eta_{\rm tot} =\eta_{\rm sys}^{(n)} 
 \otimes  \eta_{\rm env}$,  with $\eta_{\rm sys}^{(n)}= \vert n \rangle  
 \langle n\vert$, $n \in \{  1,\,\cdots, \,N_{\rm L}\}$, and 
 $ \eta_{\rm env} $, the initial density matrix
 of the environment. We define the conductance matrix 
 \begin{equation}\label{reducedcond}
 g_{nm}^{(r)} = 
 \left\langle m \left\vert  {\rm Tr}_{\rm env}\left(
 {\mathbb S} 
 \left[\eta_{\rm sys}^{(n)} \otimes \eta_{\rm env} \right]
 {\mathbb S}^{\dagger}
  \right) \right\vert m \right \rangle,
 \end{equation}
 where ${\rm Tr}_{\rm env}$ stands for the trace over the environmental
 degrees of freedom. From this matrix, 
 the dimensionless conductance is then given by,
 \begin{equation}
 g = \sum_{n=0}^{N_{\rm R} } \sum_{m=0}^{N_{\rm L}} g_{nm}^{(r)},
 \end{equation}
 and Eq.~(\ref{conductance}) reads
 \begin{eqnarray}\label{conductance_gen}
 g &=& \sum_{n \in R;m \in L} \int 
 {\rm d}{\bf q} {\rm d}{\bf q}_0 {\rm d}{\bf q}'_0 \; 
 \langle {\bf q}_0 | \eta_{\rm env} | {\bf q}'_0 \rangle
 \nonumber \\
 && \times {\mathbb S}_{m n}({\bf q}, {\bf q}_0) 
 \Big({\mathbb S}_{n m}({\bf q}, {\bf q}'_0)\Big)^*. \;
 \end{eqnarray}
 Eq.~(\ref{conductance_gen}) is the generalization 
 of the Laudauer-B\"uttiker formula in the presence of an external environment.
 It constitutes the backbone of our trajectory-based semiclassical
 theory of dephasing.

 %--------------------------------------------------------------------------------------------%
 \subsection{Drude conductance}

 The semiclassical derivation of the one particle scattering matrix has become standard~\cite{fisher_lee,Bar93,Ric-book}. Once we introduce the environment we deal with a bipartite problem, here we use  the two-particle semiclassical  propagator developed in the framework of entanglement and decoherence~\cite{Jac04, Pet06}. The extended scattering matrix elements can be written as,
  \begin{eqnarray} \label{sc-smatrix}
 {\mathbb S}_{\rm m n} (  {\bf q}, {\bf q}_{0} ) \! &=& \!  -i
  \int_{0}^{\infty}\!{\rm d}t \! \int_{\rm L}\! {\rm d}  y_{\rm 0 } \! \int_{\rm R} \!{\rm d} y
   \, {\left \langle m  \vert y  \right\rangle 
 \left\langle y_{\rm 0} \vert n \right\rangle 
 \over (2\pi \hbar)^{(d_{\rm sys}-1)/2}}
 \nonumber \\ 
 & & \! \!\! \! 
 \times 
  \sum_{\gamma,\Gamma}
 {A_{\gamma} A_{ \Gamma} \over (2\pi \hbar)^{Nd/2}}
 % \exp [
  e^{{\it i } 
  \left( 
  S_{\gamma} +
   S_{\Gamma }
    +
    {\cal S} _{\gamma,\Gamma}   \right)/\hbar}
 %    ]. 
 ,\quad \quad
 \end{eqnarray}
 where we take a $d_{\rm sys}$--dimensional, one-particle system (throughout 
 what follows we take $d_{\rm sys}=2$), and a $d$--dimensional,
 $N$-particle environment.
 At this point, ${\mathbb S}$ depends on the coordinates of the environment and is given
 by a sum over pairs of classical trajectories, 
 labeled $\gamma$ for  the system and  $\Gamma $ for the environment.  
 The classical  paths $\gamma$ and $\Gamma$ connect  
 $y_{0}$ (on a cross-section of lead L) and ${\bf q_{0}}$ (anywhere in the volume
 occupied by the environment) to $y$ (on a cross-section of lead R) and $ {\bf q }$
 (anywhere in the volume occupied by the environment) 
 in  the time $t=t_\gamma=t_\Gamma$.
 For an environment of $N$ particles in $d$ dimensions, ${\bf q}$ is a
 $Nd$ component vector.
 In the regime of pure dephasing, these paths are solely determined  by  
 $H_{\rm sys} $ and $ H_{\rm env}$. 
  Each pair of paths gives a contribution weighted by the square root $A_\gamma A_\Gamma$
 of the inverse determinant of the stability matrix~\cite{Gutzwiller,Haake-book}, and oscillating with 
 one-particle ($S_{\gamma} $ and $S_{\Gamma}$, which include Maslov indices)
 and 
 two-particle (${\cal S}_{\gamma,\Gamma} =\int_0^{t} \!{\rm d}\tau  {\cal U}[{\bf y}_{\gamma}(\tau), 
 {\bf q}_{\Gamma}(\tau)]$) action integrals accumulated along $\gamma$ and $\Gamma $. 

 We insert Eq.~(\ref{sc-smatrix}) in Eq.~(\ref{conductance_gen}), sum over channel indices with the semiclassical approximation~\cite{Jac06,Whi07} $\sum_{n}^{N_{\rm L} } \langle y_{0} \vert n \rangle  \langle n \vert  y^{\prime}_{0}  \rangle\approx \delta( y^{\prime}_{0} -y_{0})$.
 For the environment
 we make the random matrix ansatz 
 that
 $\overline{\langle {\bf q}_{0} \vert \eta_{\rm env} \vert 
 {\bf q}_{0}^{\prime} \rangle }
 \approx (2\pi\hbar)^{Nd} \; \Xi_{\rm env}^{-1}  
 \delta({\bf q}_{0}^{\prime} -{\bf q}_{0}) $,
 where $\Xi_{\rm env}$  is the environment phase-space volume. 
 The dimensionless conductance then reads
  \begin{eqnarray}\label{sc-cond}
 g &=& {(2 \pi \hbar)^{-1} \over  \Xi_{\rm env}}
  \int_{0}^{\infty}\! \! {\rm d} t  {\rm d} t^{\prime} 
  \int_{\rm env} \!\!{\rm d}  {\bf q} _{0} {\rm d} {\bf q}
  \int_{\rm L}\!\!{\rm d} y_{0}   \int_{\rm R} \!{\rm d} y
   \nonumber   \\& & \times
    \sum_{ \gamma, \Gamma ; \gamma^{\prime}, \Gamma^{\prime}}
     A_{\gamma} \; A_{\Gamma}
    \; A_{\gamma^{\prime}}\;A_{\Gamma^{\prime}} \;\;
    e^{ i (\Phi_{\rm sys} + \Phi_{\rm env} +\Phi_{\cal U})}.\qquad
 \end{eqnarray}
  This is a quadruple sum over classical paths of the system ($\gamma$ and $\gamma'$, going from $y_0$ to $y$) and the environment ($\Gamma$ and $\Gamma'$, going from ${\bf q}_0$ 
 to ${\bf q}$) with action phases,
 \begin{subequations}
  \begin{eqnarray}
  \Phi_{\rm sys} &=& \big[ 
 S_{\gamma}\left(y,y_{\rm 0}; t  \right)  - 
    S_{\gamma^{\prime}}\left(y,y_{0}; t^{\prime}  \right) \big]
 /\hbar,   
    \label{PhiS}\\
  \Phi_{\rm env} &=& \big[ 
 S_{\Gamma }\left( {\bf q},{\bf q}_{\rm 0}; t \right)-
    S_{\Gamma^{\prime}}\left( {\bf q},{\bf q}_{\rm 0}; t^{\prime}  \right) \big]
 /\hbar,    
      \label{PshiE}\\
  \Phi_{{\cal U}} &=& \big[  
  {\cal S} _{\gamma,\Gamma } (y,y_0 ;{\bf q},{\bf q}_0;t)
  -
  {\cal S} _{\gamma^{\prime},\Gamma^{\prime}}(y,y_0 ;{\bf q},{\bf q}_0;t^{\prime}) \big]
 /\hbar. 
 \nonumber \\ \label{PhiU} 
   \end{eqnarray}
 \end{subequations}

 We are interested in quantities  averaged over variations in the energy or cavity shapes. 
 For most sets of  paths the phase of a given contribution will oscillate wildly with these variations, so the contribution averages to zero. In the semiclassical limit, Eq.~(\ref{sc-cond}) is thus dominated by terms which satisfy a stationary phase condition (SPC), i.e. where the variation of
 $\Phi_{\rm sys} + \Phi_{\rm env} + \Phi_{{\cal U}}$ has to be minimized. In the
 regime of pure dephasing, individual variations of 
 $\Phi_{\rm sys}$, $\Phi_{\rm env}$ and $\Phi_{{\cal U}}$ are uncorrelated.
 They are moreover dominated by variations of $\Phi_{\rm sys}$ and $\Phi_{\rm env}$, on which
 we therefore enforce two independent SPC's.

 The dominant 
 contributions that survive averaging
 are the diagonal ones.  They give the Drude conductance. 
 Indeed setting $\gamma=\gamma^{\prime}$ and $\Gamma =\Gamma^{\prime} $
 straightforwardly satisfies SPC's over
 $\Phi_{\rm sys}$ and $\Phi_{\rm env}$.  
 These two SPC's require $t = t^{\prime}$  and lead to 
 an exact cancellation of all the phases 
 $\Phi_{\rm sys }= \Phi_{\rm env}=  \Phi_{{\cal U}} = 0$. The 
 dimensionless Drude conductance is given by
  \begin{eqnarray}
 g^{\rm D} = 
  \int_{0}^{\infty}\! \! {\rm d} t  
 \int_{\rm env} \!\! {\rm d}  {\bf q} _{0} {\rm d} {\bf q}
 \int_{\rm L}\!\!{\rm d} y_{0}   \int_{\rm R} \!\!{\rm d} y
  \sum_{\gamma,\Gamma}
 { 
 A_{\gamma}^2 \; A_{\Gamma}^2 \over (2 \pi \hbar) \, \Xi_{\rm env} } .
 \label{drudecond}
 \end{eqnarray}

 From here on, the calculation proceeds along the lines of Ref.~[\onlinecite{Jac06}].
 The main idea is to relate semiclassical amplitudes with classical probabilities. This is done by the introduction of two sum rules that express the ergodic properties of  open cavities, Eq.~(\ref{classprob_open}),  and of closed 
 ones,  Eq.~(\ref{classprob_close}),
  \begin{subequations} 
 \begin{eqnarray}  
 \hskip -6mm
 \sum_{\gamma} A_{\gamma}^2 \left[ \cdots \right]_{\gamma}  \! &=& \! \!
 \int_{-\frac{\pi}{2}}^{\frac{\pi}{2}} {\rm d} \theta_{0}{\rm d} \theta\,
 P_{\rm sys} ({\bf Y} ,{\bf Y}_0; t)
  \left[ \cdots \right]_{{\bf Y}_0}, \,
  \label{classprob_open} \\
 \hskip -6mm
 \sum_{\Gamma} A_{\Gamma}^2 \left[ \cdots \right]_{\Gamma} \!
 &=&  \! \!
 \int  {\rm d} {\bf p}_0 \,{\rm d} {\bf p}\, 
 \tilde{P}_{\rm env} ({\bf Q} ,{\bf Q}_0 ; t )
 \left[ \cdots \right]_{{\bf Q}_0}.\,
 \label{classprob_close}
 \end{eqnarray}
 \end{subequations}
 Here, $P_{\rm sys} ({\bf Y},{\bf Y}_0; t)  = 
 p_{\rm F} \cos \theta_0 \times \tilde{P}_{\rm sys} ({\bf Y} ;{\bf Y}_0; t)$,
 and 
 $ \tilde{P}_{\rm sys} ({\bf Y},{\bf Y}_0; t)$  and 
 $\tilde{P}_{\rm env} ({\bf Q},{\bf Q}_0; t)$ are the  classical probability densities. For the system,
 we need to take into account the fact that particles are injected,
 which is why the classical probability density must be multiplied with
 the initial system momentum $p_{\rm F} \cos \theta_0 $
 along the injection lead \cite{Bar93}.  
 The phase points  ${\bf Y}_{0}= \left( y_{0},  \theta_0\right) $ and ${\bf Y} =   \left( y,  \theta \right) $ are at the boundary between
 the system and the leads. 
 In contrast ${ \bf Q}_{0 }=\left( {\bf q}_{0},  {\bf p}_{0}\right) $   
 and ${\bf Q} = \left( {\bf q},  {\bf p}  \right) $ are inside the closed 
 environment cavity. The momenta are integrated over the
 entire environment phase-space, while 
 $\tilde{P}_{\rm env} ({\bf Q},{\bf Q}_0; t)$ will always contain a
 $\delta$-function which restricts the final energy to equal the initial
 one, (i.e.~$|{\bf p}|=|{\bf p}_0|$ 
 if all $N$ environment particles have the same mass).

 The average of $P_{\rm sys} $ over an ensemble of systems or over energy gives 
 a smooth function.  For a chaotic system we write  
  \begin{subequations}
 \begin{eqnarray}
 &&
 \left \langle \tilde{P}_{\rm sys} ({\bf Y};{\bf Y}_0; t) \right\rangle =
 \frac{ \cos \theta }{2 \left( W_{\rm L} + W_{\rm R} \right) \tau_{\rm D} } 
 e^{-t/\tau_{\rm D} } \; . \qquad \label{classprobaveS} 
 \end{eqnarray}
 Likewise, the average of $\tilde{P}_{\rm env}$ gives
 \begin{eqnarray}
 &&
 \left \langle\tilde{P}_{\rm env} ({\bf Q};{\bf Q}_0; t^{\prime}) \right\rangle =
 {\delta (|{\bf p}|-|{\bf p}_0|) \over \Xi_{\rm env}^{|{\bf p}_0|} }  \, ,
 \label{classprobaveE}
 \end{eqnarray}
 \end{subequations}
 where $\Xi_{\rm env}^{|{\bf p}_0|}$ is the size of the hypersurface in
 the environment's
 phase-space defined by $|{\bf p}|=|{\bf p}_0|$
 (for $d=2$, $\Xi_{\rm env}^{|{\bf p}_0|} =(2\pi p_{\rm F}^{\rm env}\Omega_{\rm env})^N$
 where $\Omega_{\rm env}$ is the area of the environment).
 Inserting Eqs.~(\ref{classprob_open}),
 (\ref{classprob_close}), (\ref{classprobaveS}) and~(\ref{classprobaveE})
 into Eq.~(\ref{drudecond}),
 we can perform all integrals 
 using 
 $\int_{\rm env} \rmd {\bf Q}_0 
 \equiv \int_{\rm env} \rmd {\bf q}_0\rmd {\bf p}_0 = \Xi_{\rm env}$ 
 and 
 $\int_{\rm env} \rmd {\bf Q} \delta (|{\bf p}|-|{\bf p}_0|) 
 = \Xi_{\rm env}^{|{\bf p}_0|}$.
 Then, since $ N_{\rm L, R} = k_{\rm F} W_{\rm L, R} /\pi$, 
 we recover the classical Drude conductance,
 \begin{equation}\label{eq:Drudecond}
 g^{\rm D} = \frac{ N_{\rm L}N_{\rm R}} {N_{\rm L} +N_{\rm R}}.
 \end{equation}

  %--------------------------------------------------------------------------------------------%
 \subsection{Overview of the effect of environment on
 weak-localization}
 \label{sect:weak-loc-env}

\begin{figure}
 \includegraphics[width=7cm]{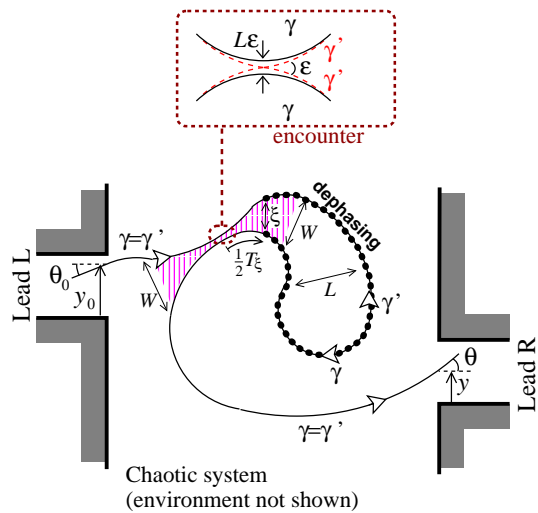} 
 \caption{\label{fig:weakloc} 
 (color online) A semiclassical contribution to weak-localization for 
 the system-environment model.
 The paths are paired everywhere except at the 
 encounter, where one path crosses itself at angle $\epsilon$,
 while the other one does not (going the opposite way around the loop).
 Here we show $\xi>\epsilon L$, so the dephasing (dotted path segment)
 starts in  the loop ($T_{\xi} >0$). 
 }
 \end{figure}

 The leading-order weak-localization corrections to the conductance
 were identified in Refs.~[\onlinecite{Ale96,Sie01,Ric02}]
 as those arising from trajectories that are paired almost everywhere 
 except in the vicinity of an encounter. An example of such a trajectory 
 is shown in Fig.~\ref{fig:weakloc}.  
 At the encounter, one of the  trajectories ($\gamma'$) intersects itself, 
 while the other one ($\gamma$) avoids the crossing. 
 Thus, they travel along the loop they form in opposite directions. 
 For chaotic ballistic systems 
 in the semiclassical limit, only pairs of trajectories with
 small crossing angle $\epsilon$ contribute significantly to weak-localization.
 In this case, trajectories remain correlated for some time
 on both sides of the encounter, the correlated region indicated in pink in 
 Fig.~\ref{fig:weakloc}. In other words, the smallness of $\epsilon$
 requires two minimal times, $T_{\rm L}(\epsilon)$ to form a loop,
 and $T_{\rm W}(\epsilon)$ in order for the legs to separate
 before escaping into different leads.
 In the case
 of an hyperbolic dynamics one estimates~\cite{caveat-time},
 \begin{subequations}
   \begin{eqnarray}\label{TL}
 T_{\rm L}(\epsilon) &  \simeq  & \lambda^{-1} \ln [\epsilon^{-2}], \\
 \label{TW}
 T_{\rm W}(\epsilon) & \simeq & \lambda^{-1} \ln [\epsilon^{-2} (W/L)^2].
 \end{eqnarray}
 \end{subequations}

 As long as the system-environment coupling does not generate
 energy and/or momentum relaxation, the presence of an environment
 does not significantly change 
 this picture. However it does lead to dephasing via the accumulation of
 uncorrelated action phases, mostly along the loop, 
 when $\gamma$ and $\gamma'$ are
 more than a distance $\xi$ apart. This is illustrated in
 Fig.~\ref{fig:weakloc} for the case when $\xi$ is less than $W$.
 We define a new timescale $T_{\xi}$ as twice the time 
 between the encounter and the start of the dephasing,
 \begin{equation}\label{Txi}
 T_{\xi}(\eps)  \approx  \lambda^{-1} \ln [\epsilon^{-2}(\xi/L)^2].
 \end{equation} 
 Dephasing occurs mostly in the loop part. However if $\xi < \epsilon L$
 and $T_{\xi}<0$,
 dephasing starts before the paths reach the encounter.  
 We discuss this point in more detail below in Section~\ref{env-discus}.

 %%%%%%%%%%%%%%%%%%%%%%%%%%%%%%%%%%%%%%%%% %%%%%%%
 %%%%%%%%%%%%%%%%%%%%%%%%%%%%%%%%%%%%%%%%%%%%%%%%

 \subsection{Calculating the effect of the environment on weak-localization}
 \label{sect:weak-loc-calc}

 In the absence of dephasing each weak-localization contribution accumulates 
 a phase difference
 $\delta \Phi_{\rm sys} = E_{\rm F} \epsilon^2/(\lambda\hbar)$~\cite{Ric02,Sie01}.  
 In the presence of an environment, an additional  action phase 
 difference $\delta\Phi_{\cal U}$ is accumulated. 
 Incorporating this additional phase into
 the calculation of weak-localization does not require significant departure
 from the theory at ${\cal U}=0$. We extend the theory of Ref.~[\onlinecite{Jac06}] to account for this additional phase.

 We follow the same route as for the Drude conductance,
 but now consider the pairs of paths described in 
 Section~\ref{sect:weak-loc-env} above, and shown
 in Fig.~\ref{fig:weakloc}, while the environment is still 
 treated within the diagonal approximation, $\Gamma'=\Gamma$.
 The sum rule of Eq.~(\ref{classprob_close}) still applies.
 Though the sum over system paths is restricted to paths with an encounter,
 we can still write this sum in the form given in Eq.~(\ref{classprob_open}),
 provided the probability $P_{\rm sys} ({\bf Y},{\bf Y}_0;t)$ is
 restricted to paths which cross themselves.  
 To ensure this we write
 \begin{eqnarray}
 \!\!P_{\rm sys} ({\bf Y},{\bf Y}_0;t)
 \!&=& \!\! p_{\rm F} \cos \theta_0 
 \int_{C} {\rm d} {\bf R}_2 {\rm d} {\bf R}_1
 \tilde{P}_{\rm sys} ({\bf Y},{\bf R}_2;t-t_2)
 \nonumber \\
 \!& \times & \!\!
 \tilde{P}_{\rm sys} ({\bf R}_2,{\bf R}_1;t_2-t_1)
 \tilde{P}_{\rm sys}({\bf R}_1,{\bf Y}_0;t_1) \,. \qquad
 \end{eqnarray} 
 Here, we use ${\bf R}=({\bf r},\phi)$, $\phi \in [-\pi,\pi]$ for phase-space
 points inside the cavity, while ${\bf Y}$ lies on the lead
 as before. We then restrict the 
 probabilities inside the integral to  trajectories which cross themselves at 
 phase-space positions ${\bf R}_{1,2}$ with the first (second) visit to the 
 crossing occurring at time $t_1$ ($t_2$).  We can write
 ${\rm d} {\bf R}_2 = v_{\rm F}^2 \sin \epsilon {\rm d}t_1 {\rm d}t_2{\rm d} \epsilon$ and set 
 ${\bf R}_2 =({\bf r_1},\phi_1\pm \epsilon)$.  
 Then, the weak-localization correction to the dimensionless
 conductance 
 in the presence of an environment is given by,
 \begin{subequations}
 \label{eq:gwl-integral}
 \begin{eqnarray}
 g^{\rm wl} &=&  (\pi \hbar)^{-1} \int_{\rm L} \!  {\rm d} {\bf Y}_{0}\int {\rm d} \epsilon \,
 {\rm Re}\big[e^{{\it i} \delta \Phi_{\rm sys} }\big] 
 \big\langle  F ( {\bf Y}_{0}, \epsilon)  \big\rangle, \qquad 
 \label{eq:gwl-integral-a}
  \end{eqnarray}
  \noindent with,
 \begin{eqnarray}\label{eq:F}
  F( {\bf Y}_{0}, \epsilon)  &=&
 2v_{\rm F}^2  \sin \epsilon
  \int_{T_{\rm L}+T_{\rm W}} ^{\infty} \!  {\rm d } t 
 \int_{T_{\rm L}+\frac{T_{\rm W}}{2}}^{t-\frac{T_{\rm W} }{2} } \!   {\rm d} t_2
 \int_{\frac{T_{\rm W}}{2}}^{t_2-T_{\rm L}}  \!   {\rm d} t_1
 \nonumber \\ &\times&
 p_{\rm F}\cos \theta_0 \int_{\rm R} \!  {\rm d} {\bf Y}
 \int_{ C} \! {\rm d} {\bf R}_1 
 \tilde{P}_{\rm sys}({\bf Y} ,{\bf R}_2 ;t-t_2)  
 \nonumber \\ 
 &\times&
 \tilde{P}_{\rm sys}({\bf R}_2,{\bf R}_1;t_2-t_1)  
 {\tilde P}_{\rm sys}({\bf R}_1,{\bf Y}_0;t_1)
 \nonumber \\
 &\times& 
 \int{\rmd {\bf Q} \rmd{\bf Q}_0 \over \Xi_{\rm env}}
 {\tilde P}_{\rm env}({\bf Q},{\bf Q}_0;t)
 \exp[{\rmi \delta \Phi_{\cal U}}] .  
 \end{eqnarray}
 Comparison with Eq.~(34) of Ref.[\onlinecite{Jac06}] shows that the 
 effect of the environment is entirely contained in the last line of
 Eq.~(\ref{eq:F}).
 At the level of the diagonal approximation for the environment,
 $\Gamma'=\Gamma$, one has
 \begin{eqnarray}
 \delta \Phi_{\cal U}= {1\over \hbar} \! \int_0^t \rmd \tau 
 \big[{\cal U}\big({\bf r}_\gamma(\tau),{\bf q}_\Gamma(\tau))
 - {\cal U}\big({\bf r}_{\gamma'}(\tau),{\bf q}_\Gamma(\tau) \big)\big],
 \nonumber \\
 \end{eqnarray} 
 \end{subequations}
 where ${\bf r}_\gamma(\tau)$ and ${\bf q}_\Gamma(\tau)$ parametrize
 the trajectories of the system and of the environment respectively.
 We note that in the absence of coupling, $\delta \Phi_{\cal U} = 0 $, 
 the integral over the environment is one, 
 and we recover the weak-localization correction of an isolated system 
 (cf.~Eq.~$(35)$ of Ref.~[\onlinecite{Jac06}]).  

 To evaluate Eqs.~(\ref{eq:gwl-integral}), we need the 
 average effect of the environment on the system,
 after one has traced out the environment. For a single measurement,
 this average is an integral over all classical paths followed by 
 the environment, starting from its initial state at the beginning of the
 measurement. We therefore also average over an ensemble of initial 
 environment states, or an ensemble of environment Hamiltonians,
 which corresponds to performing many measurements.
 For compactness we define $\langle\cdots \rangle_{\rm env}$ 
 as this integral over environment paths and the 
 ensemble averaging over the environment,
 \begin{eqnarray}
 \label{eq:env-average}
 \langle\cdots \rangle_{\rm env}
 = \int {\rmd {\bf Q}  \, {\rm d} {\bf Q}_{0} \over \Xi_{\rm env}}
 \,
 \big\langle \tilde{P}_{\rm env} ({\bf Q} ;{\bf Q}_0 ; t )
 \left[ \cdots \right]_{{\bf Q}_0}  \big \rangle \,.
 \end{eqnarray}
 Without loss of generality we assume that for all ${\bf r}$,
 the interaction ${\cal U}({\bf r},{\bf q})$ is zero upon averaging 
 over all ${\bf q}$. We can ensure an 
 arbitrary interaction fulfills this condition by moving
 any constant term in ${\cal U}$ into the system Hamiltonian (these terms
 do not lead to dephasing). Since the environment is ergodic, we have   
 \begin{eqnarray}
 \big\langle {\cal U}\big({\bf r}_\gamma(t),{\bf q}_\Gamma(t)\big) 
 \big\rangle_{\rm env}
  =0 \,.
 \end{eqnarray}
 Now we use the chaotic nature of the environment to give the
 properties of the correlation function 
 $\big\langle {\cal U}\big({\bf r}_\gamma(t),{\bf q}_\Gamma(t)\big)
 {\cal U}\big({\bf r}_{\gamma'}(t'),{\bf q}_\Gamma(t')\big) 
 \big\rangle_{\rm env}$.
 \begin{itemize}
 \item[(i)] Correlation functions typically decay exponentially fast
 with time in chaotic systems, with a typical decay time
 related to the Lyapunov exponent~\cite{Eck04}. 
 The precise functional 
 form $J(\lambda_{\rm env}|t'-t|)$ of the 
 temporal decay of the coupling correlator
 depends on details of $H_{\rm env}$ and ${\cal U}$,
 however for all practical purposes, it is sufficient to know
 that it decays fast, and we approximate it by a $\delta$-function,
 $J(\lambda_{\rm env}|t'-t|) \simeq \lambda_{\rm env}^{-1} \delta(t)$.

 \item[(ii)]
 We argue that the spatial correlations 
 of ${\cal U}$ for two different system paths
 at the same $t$ also decay,
 because the averaging over many paths, and many initial environment
 states act like an average over ${\bf q}$.  
 We define $K(|{\bf r}'-{\bf r}|/\xi)$ as the functional form of this 
 decay of spatial correlations, with $K(0)=1$. 
 The precise form of $K(x)$  
 depends on details of $H_{\rm sys}$,
 $H_{\rm env}$ and ${\cal U}$.
 In particular, the typical length $\xi$ of this decay 
 is of order the scale on which 
 ${\cal U}({\bf r},{\bf q})$ changes between its maximum and minimum value.
 \end{itemize}
 Given these arguments, we have
 \begin{eqnarray}
 & & \hskip -8mm
 \big\langle {\cal U}\big({\bf r}_\gamma(t),{\bf q}_\Gamma(t)\big)
 {\cal U}\big({\bf r}_{\gamma'}(t'),{\bf q}_\Gamma(t')\big) 
 \big\rangle_{\rm env}
 \nonumber 
 \\
 &=& {\langle {\cal U}^2 \rangle \over \lambda_{\rm env}}
 \ K \left({ |{\bf r}_{\gamma'}(t)-{\bf r}_{\gamma}(t)|/\xi}\right) \;
 \delta(t-t') \, .
 \label{eq:U-correl}
 \end{eqnarray}
 Then,
 \begin{eqnarray}
 & & \hskip -5mm
 \langle \de \Phi_{\cal U}^2 \rangle_{\rm env}
 \nonumber \\
 &=& {1 \over \hbar^2}\int_0^t \rmd \tau_2 \rmd \tau_1 \nonumber \\
 & & \qquad\times
 \Big\langle 
 \big[
 {\cal U}\big({\bf r}_\gamma(\tau_2),{\bf q}_\Gamma(\tau_2)\big)-
 {\cal U}\big({\bf r}_{\gamma'}(\tau_2),{\bf q}_\Gamma(\tau_2)\big) 
 \big] 
 \nonumber \\
 & & \qquad\times
 \big[
 {\cal U}\big({\bf r}_\gamma(\tau_1),{\bf q}_\Gamma(\tau_1)\big)-
 {\cal U}\big({\bf r}_{\gamma'}(\tau_1),{\bf q}_\Gamma(\tau_1)\big) 
 \big] 
 \Big\rangle_{\rm env}
 \nonumber \\
 &=& 2 {\langle {\cal U}^2 \rangle \over \lambda_{\rm env}}
 \int_0^t \rmd \tau
 \big[1 - K(|{\bf r}_{\gamma'}(\tau)-{\bf r}_\gamma(\tau)|/\xi)\big] \,.
 \end{eqnarray}

 We further make the following step-function approximation for $K$,
 \begin{eqnarray}
 K(x) = \Theta(1-x),
 \label{eq:f-approx}
 \end{eqnarray}
 where the Euler $\Theta$-function is one (zero) 
 for positive (negative) arguments.  
 In principle, this is unjustifiable for $x \sim 1$,
 however since the paths diverge exponentially from
 each other, the time during which $x\sim 1$ is of order $\lambda^{-1}$, 
 while dephasing happens on a timescale $\tau_\phi$ which is typically 
 of order the dwell time, $\tau_{\rm D}$.
 Thus the step-function approximation of $K(x)$ will have corrections
 of order $(\lambda\tau_{\rm D})^{-1} \ll 1$, which we therefore
 neglect.
 Once we have made the approximation in Eq.~(\ref{eq:f-approx}),
 we see that non-zero contributions to
 $\langle \de \Phi_{\cal U}^2 \rangle$ come from regions
 where the distance between $\gamma$ and $\gamma'$ is 
 larger than $\xi$.

 We are now ready to calculate dephasing for those 
 system paths shown in Fig.~\ref{fig:weakloc}.
 As defined above, $t_1$ and $t_2$ are the two times at which 
 the path $\gamma'$ crosses itself. 
 Dephasing acts on the loop formed by $\gamma'$ and, as just argued,
 it acts once the distance between $\gamma$ and $\gamma'$ is 
 greater than $\xi$, i.e. in the time window from
 $(t_1 + T_\xi/2)$ to $(t_2-T_\xi/2)$, where 
 $T_\xi(\eps)$ is given in Eq.~(\ref{Txi}).
 We average Eq.~(\ref{eq:F}) over the environment
 and use the central limit theorem to evaluate 
 the action phase due to the coupling between system and
 environment,
 \begin{eqnarray}
 \big\langle \exp[{\rmi \delta \Phi_{\cal U }}] \big\rangle_{\rm env} 
 &=&
  \exp \Big[-\half \langle\delta \Phi^2_{\cal U}\rangle_{\rm env} \Big]
 \nonumber \\
 &=&
  \exp\left[ - (t_2-t_1-T_{\xi})/\tau_{\phi}\right],
 \label{detphiu}
 \end{eqnarray}
 where the dephasing rate is 
 \begin{eqnarray}\label{dephi_rate}
 \tau_{\phi}^{-1} \sim
 \hbar^{-2} \lambda_{\rm env}^{-1}
 \langle {\cal U}^2 \rangle.
 \end{eqnarray}

 Given that $\langle \cdots \rangle_{\rm env}$ is defined
 in Eq.~(\ref{eq:env-average}), we can substitute
 Eq.~(\ref{detphiu}) directly into Eqs.~(\ref{eq:gwl-integral}).
 We thereby
 reduce the problem to an integral over system paths
 which is almost identical to the equivalent integral for 
 ${\cal U}=0$. 
 Assuming phase space ergodicity  for the system, we get 
 the probabilities
 \begin{subequations}
 \begin{eqnarray}
 \langle \tilde{P}_{\rm sys} ({\bf R}_1,{\bf Y}_{0};t_1)\rangle   
 & = & {e^{-t_1/\tau_{\rm D}} \over 2\pi \Omega_{\rm sys }}  \, , 
 \\   
 \langle \tilde{P}_{\rm sys}({\bf R}_2,{\bf R}_1;t_2-t_1)\rangle  &=& 
 {e^{-(t_2-t_1-T_{\rm W}/2)/\tau_{\rm D}} \over
 2\pi \Omega_{\rm sys}} \, ,
 \\
 \langle \tilde{P}_{\rm sys}({\bf Y},{\bf R}_2 ;t-t_2)\rangle   
 &=& {\cos \theta  \, e^{-(t-t_2-T_{\rm W}/2)/\tau_{\rm D}}
 \over 2(W_{\rm L}+W_{\rm R})\tau_{\rm D}} , \qquad
 \end{eqnarray}
 \end{subequations}
 with $\Omega_{\rm sys} $ being the real space volume occupied by
 the system (the area of the cavity). At this point 
 the integral is the same as without dephasing, 
 except that during the time $(t_2-t_1-T_{\xi})$, 
 the inverse dwell time is replaced by $(\tau_{\rm D}^{-1}+\tau_\phi^{-1})$.
 Thus when evaluating the $(t_2-t_1)$-integral, 
 we get the extra prefactor 
 $\exp\left[ - T_{\rm L}(\eps)/\tau_{\phi} \right]/(1+ \tau_{\rm D}/\tau_{\phi})$
 compared with the equivalent integral result without dephasing.
 Thus we have
 \begin{eqnarray}\label{eq:Fres}
 \left\langle  F( {\bf Y}_{0}, \epsilon)  \right\rangle
 &\propto & \sin \eps\, 
 \frac{\; e^{-T_{\rm L}(\eps)/\tau_{\rm D}
 - (T_{\rm L}(\eps)-T_{\xi}(\eps))/\tau_{\phi}}}{1+ \tau_{\rm D}/\tau_{\phi}}.\quad 
 \end{eqnarray}
 Since $(T_{\rm L}(\eps)-T_{\xi}(\eps))=\tau_\xi$
 with $\tau_\xi$ defined in Eq.~(\ref{eq:tau_xi}),
 the $\eps$ dependence of the $\tau_{\phi}^{-1}$-term drops out.
 This means that $\left\langle  F( {\bf Y}_{0}, \epsilon) \right\rangle$ 
 simply differs from its value without dephasing by a constant factor,
 $\e^{-\tau_\xi/\tau_\phi}/(1+ \tau_{\rm D}/\tau_{\phi})$. 
 Thus the integral over $\eps$ in Eq.~(\ref{eq:gwl-integral-a}) 
 is identical to the one in the absence of the 
 environment, and 
 takes the form\cite{Jac06} ${\rm Re} \int_0^\infty \rmd \eps \,
 \eps^{1+ 2/(\lambda \tau_{\rm D})}
 \,\exp[\rmi E_{\rm F} \eps^2/(\lambda \hbar)]$, where we have assumed
 $\eps \ll 1$.
 The substitution $z=E_{\rm F} \eps^2/(\lambda \hbar)$ immediately 
 yields a dimensionless integral
 and an exponential term, $\e^{-\tau_{\rm E}^{\rm cl}/\tau_{\rm D}}$
 (neglecting as usual ${\cal O}[1]$-terms in the logarithm in 
 $\tau_{\rm E}^{\rm cl}$).
 From this analysis, we find that the weak-localization correction
 is given by
 \begin{equation}\label{eq:gwl-env}
 g^{\rm wl} = 
 \frac{g^{\rm wl}_0 }{1+\tau_{\rm D}/\tau_\phi}
  \; \exp[-\tau_{\xi}/\tau_\phi],
 \end{equation}
 where
 $g_0^{\rm wl}$ is the weak-localization correction at 
 finite-$\tEc$ in the absence of dephasing,
 \begin{eqnarray}\label{eq:gwl0}
 g_0^{\rm wl} = - {N_{\rm L} N_{\rm R}\over(N_{\rm L} +N_{\rm R})^2}
 \exp[-\tau_{\rm E}^{\rm cl} / \tau_{\rm D}].
 \end{eqnarray}
 We see that the dephasing of weak localization
 is not exponential with the Ehrenfest time,
 instead it is exponential with the $\lambda_{\rm F}$-independent
 scale $\tau_\xi$ 
 given in Eq.~(\ref{eq:tau_xi}).  
 In all cases where $\xi$ is a classical scale
 (i.e. of similar magnitude to $W,L$ rather than $\lambda_{\rm F}$)
 we see that $\tau_\xi$ is much less than the Ehrenfest time,
 $\tau_{\xi} \ll \tau_\phi$.  In such cases the 
 exponential term in Eq.~(\ref{eq:gwl-env}) is  much less noticeable that the universal
 power-law suppression of weak-localization.

 %--------------------------------------------------------------------------------------------%
 \subsection{Weak-localization for reflection and coherent backscattering}

\begin{figure}
 \includegraphics[width=7cm]{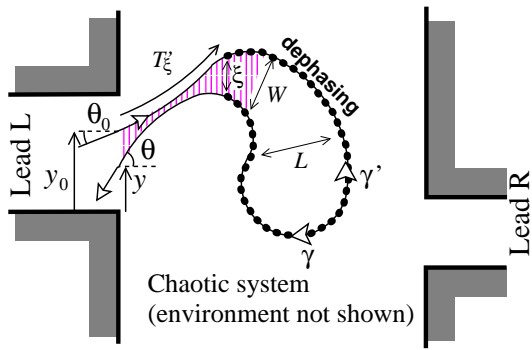} 
 \caption{\label{fig:cbs} 
 (color online) A semiclassical contribution to coherent backscattering for 
 the system-environment model.
 It involves paths which return to close, but 
 anti-parallel to themselves at lead $L$. The two solid paths are
 paired (within $W$ of each other) in the
 cross-hatched region.
 Here we show $\xi>\epsilon L$, so the dephasing (dotted path segment)
 starts in  the loop ($T_{\xi} >0$).  
 In the basis parallel and perpendicular to $\gamma$ at injection the initial position and  momentum of path $\gamma$ at exit are  $r_{0\perp} = (y_{0} - y)\cos\theta_0$,  $r_{0\pll} = (y_{0} - y)\sin\theta_0$ and  $p_{0\perp} = p_{\rm F} (\theta -\theta_0) $.
 }
 \end{figure}

 We show explicitly that our semiclassical method is probability-conserving,
 and thus current-conserving, also in the presence of dephasing.
 We do this by calculating the leading-order quantum corrections to reflection,
 showing that they enhance reflection by exactly the
 same amount that transmission is reduced.
 There are two leading-order off-diagonal corrections to reflection.
 The first one reduces the probability of reflection to arbitrary momenta  
 (weak-localization for reflection), 
 while the second one enhances the probability of 
 reflection to the time-reversed 
 of the injection path (coherent-backscattering).
 The distinction between these two contributions is related to the correlation 
 between the path segments when they hit the leads. 
 For coherent-backscattering contributions, these segments
 are correlated (see Fig.~\ref{fig:cbs}),
 but for weak-localization contributions, they are not.

 The derivation of the weak-localization for reflection  $r^{\rm wl} $  
 is straightforward and proceeds in the same way as the derivation
 for $g^{\rm wl}$ given above, replacing the factor 
 $N_{\rm R} / (N_{\rm R}  + N_{\rm L})$ by  
 $N_{\rm L} / (N_{\rm R}  + N_{\rm L})$. We thus get, 
 \begin{equation}\label{eq:rwl-env}
 r^{\rm wl} = 
 \frac{r^{\rm wl}_0 }{1+\tau_{\rm D}/\tau_\phi}
  \; \exp[-\tau_{\xi}/\tau_\phi],
 \end{equation}
 where $r_0^{\rm wl} = - \exp[-\tau_{\rm E}^{\rm cl} / \tau_{\rm D}] \;
 N_{\rm L}^2 /(N_{\rm L} +N_{\rm R})^2$ is the finite-$\tEc$ correction 
 in the absence of dephasing.

 %%%%%%%%%%%%%%%%%%%%%%%%%%%%%%%%%%%%%%%%% %%%%%%%
 %%%%%%%%%%%%%%%%%%%%%%%%%%%%%%%%%%%%%%%%% %%%%%%%

 We next calculate the contributions to coherent-backscattering,
 extending the treatment of Ref.~[\onlinecite{Jac06}] to account for
 the presence of dephasing. As before, 
 the environment is treated in the diagonal approximation.  
 The coherent-backscattering contributions 
 correspond to trajectories where legs escape together 
 within $T_{\rm W}/ 2$ of the encounter.
 Such a contribution is shown in  Fig.~\ref{fig:cbs}. 
 The correlation between the system paths at injection and exit 
 induces an action difference $\delta   \Phi_{\rm sys} = \delta S_{cbs}$ 
 not given by the Richter-Sieber  expression.  
 It is convenient to write this action difference 
 in terms of relative coordinates at the lead (rather than at the encounter).
 The system action difference is then
 $ \delta S_{cbs} = -( p_{0\perp} + m \lambda r_{0\perp}) r_{0\perp} $
 where the perpendicular difference in position and momentum are
 $r_{0\perp} = (y_{0} - y)\cos\theta_0$ and  
 $p_{0\perp} = p_{\rm F} (\theta -\theta_0)$.
 As with weak-localization, we can identify three timescales,
 $T'_{\rm L},T'_{\rm W},T'_\xi$,
 associated with the time for paths to spread to each of three length scales,
 $L,W,\xi$.  However unlike for weak-localization we define these timescales
 as a time measured from the lead rather than from the encounter.
 Thus we have
 \begin{eqnarray}
 \label{eq:Tprimed}
 T^{\prime}_{\ell }(r_{0\perp}, p_{0\perp}) \simeq \lambda^{-1} 
 \ln [(m\lambda  {\ell})^2 / \vert p_{0\perp} + m\lambda r_{0\perp}  \vert^2] 
 \end{eqnarray}
 with ${\ell}  =\{ L,W,  \xi \}$.
 Writing the integral over ${\bf Y}_0$ as an integral over 
 $(r_{0\perp},p_{0\perp})$ and using 
 $p_{\rm F}  \sin \theta_0 {\rm d} {\bf Y}_{0} =  {\rm d}p_{0\perp}{\rm d}r_{0\perp}$,
 the coherent-backscattering contribution is 
 \begin{eqnarray}
 r^{\rm cbs} &=&  \int_{\rm L} \!  
 {{\rm d}p_{0\perp}{\rm d}r_{0\perp} \over p_{\rm F}  \sin \theta_0}
 {\rm Re}\big[e^{{\it i} \delta S_{\rm cbs} }\big] 
 \big\langle  F^{\rm cbs} ({\bf Y}_{0})  \big\rangle \,. \qquad 
  \end{eqnarray}
 After integrating out the environment 
 in the same manner as for weak-localization, we get
 \begin{eqnarray}
   \label{eg:Fcbs}
  && \hskip -6mm
 F^{\rm cbs} ({\bf Y}_{0})  
 \\
 &=&
  \int_{\rm L} \!  {\rm d} {\bf Y} \int_{T'_{\rm L}}^{\infty} {\rm d}t  
  \langle P_{\rm sys } ({\bf Y},{\bf Y}_0, t) \rangle  
 \exp [-(t-T'_\xi)/\tau_\phi]
   \nonumber\\
 &=&
  \frac{N_{\rm L}p_{\rm F}  \sin \theta_0 }{\pi(N_{\rm L}+N_{\rm R})}  
  {\exp\big[- (T'_{\rm L} -T'_W/2)/\tau_{\rm D}
 - (T'_{\rm L}-T'_\xi)/\tau_\phi
 \big] \over 1+ \tau_{\rm D} /\tau_\phi}\, . \ \ \ 
 \nonumber
  \end{eqnarray}

 Now we can proceed as for ${\cal U}=0$, pushing the momentum integral's 
limits to infinity,  and  evaluating  the $r_{0\perp}-$integral 
 over the range $W$,
 with the help of an Euler $\Gamma$-function. We finally obtain
  \begin{equation}\label{eq:rcbs-env}
 r^{\rm cbs} = 
 \frac{r^{\rm cbs}_0 }{1+\tau_{\rm D}/\tau_\phi}
  \; \exp[-\tau_{\xi}/\tau_\phi],
 \end{equation}
 in terms of 
 $r_0^{\rm cbs} =  \exp[-\tau_{\rm E}^{\rm cl} / \tau_{\rm D}] \;
 N_{\rm L} /(N_{\rm L} +N_{\rm R})$,
 the finite-$\tEc$ coherent backscattering contribution
 in the absence of dephasing.
 Hence  $r^{\rm cbs} + r^{\rm wl} = - g^{\rm wl}$ for all values of
 $\tau_\xi$ and $\tau_\phi$, and 
 our approach  is  probability- and thus current-conserving.

 \subsection{Weak-localization corrections in the environment}

 So far, we have only considered cases where
 we make a diagonal approximation for the environment.
 On the face of it this seems a little unreasonable. For instance,
 if the system and environment are of similar sizes
 then one would expect that diagonal contributions for the system 
 and weak-localization for the environment
 would be as important as the contributions calculated above.

 Since the environment is a closed cavity, one would naively think that
 the weak-localization contribution for the environment should be calculated
 in a similar manner to the form-factor in Refs.~[\onlinecite{Sie01,Berko02}].
 In the absence of coupling, the environment part of Eq.~(\ref{sc-cond}) 
 would then be
 \begin{eqnarray}
 \int \! {\rmd {\bf q} \rmd {\bf q}_0 \over \Xi_{\rm env}}
 \sum_{\Gamma,\Gamma'} A_\Gamma A_{\Gamma'} 
 \e^{\rmi \Phi_{\rm env}}
 = 1 + \mu {\Delta_{\rm env} t \over \hbar}, 
 \label{eq:env-weak-loc}
 \end{eqnarray}
 where $\mu$ is a number of order one, and 
 $\Delta_{\rm env}$
 is the environment level-spacing
 (for a two-dimensional environment containing a single particle,  
 $\Delta_{\rm env} \sim  \hbar^2 /mL_{\rm env}^2$).
 The first term above comes from the diagonal approximation used throughout
 this article, while the second term is a weak-localization correction.
 This correction becomes 
 of order the system's weak-localization correction 
 on the timescale $t\sim \tau_{\rm D}$, so there is a priori no reason
 to neglect it.  

 The environment part of our calculation differs however from the form-factor
 in that it corresponds to the time-evolution of the environment
 during the time it takes for a particle to be transported across
 the system. Therefore, the sum in Eq.~(\ref{eq:env-weak-loc})
 is not restricted to periodic orbits, and the unitarity of the 
 environment's time-evolution imposes that $\mu=0$. 
 Furthermore, unitarity must be preserved even in the 
 presence of a finite-${\cal U}$, as long
 as there is no exchange of particles between system and environment.
 We thus conclude that we do not need to consider the weak-localization
 type corrections to the environment evolution because they cancel.

 %---------------------------------------------------------------------%
 \subsection{Regime of validity of the semiclassical calculation}
 \label{env-discus}

 Throughout this article we assumed that the system-environment
 coupling is weak enough not to modify the 
 classical paths in the system. Formally, this assumption can be
 rigorously justified by invoking theorems on structural 
 stability~\cite{Hirsch}. However,  
 care should be taken in extrapolating our results to the limit 
 $\xi \rightarrow 0$, since the force on the particle is the
 gradient of the interaction potential, $\sim U/\xi$.
 We therefore estimate the minimum $\xi$ for which 
 we can legitimately assume that classical system paths are left
 unchanged by the system-environment coupling.
 This will give the bound on the regime of validity of our
 approach.

 %%%%%%%%%%%%%%%%%%%%%%%%%%%%%%%%%%%%%%%%% %%%%%%%
 \begin{figure}
 \begin{center}
 \includegraphics[width=7cm]{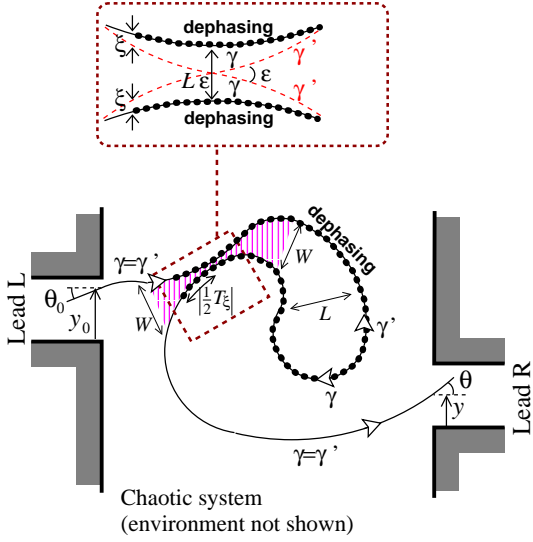}
 \caption{\label{fig:negative-tau_xi}
 (color online) 
 Dephasing of weak-localization when  
 $\xi \ll L\eps\sim (L\lambda_{\rm F})^{1/2}$,
 and hence $T_\xi(\eps)$ is negative, see Eq.~(\ref{Txi}).
 The dephasing starts and ends in the ``legs'' rather than the loop.
 In the language of disordered systems this means the dephasing
 affects the diffusons as well as the cooperon.
 }
 \end{center}
 \end{figure}
 %%%%%%%%%%%%%%%%%%%%%%%%%%%%%%%%%%%%%%%%% %%%%%%%

 To see significant dephasing we need $\tau_\phi \sim \tau_D$,  
 so we cannot take the interaction strength to zero, instead we require
 that 
 $\langle {\cal U}^2 \rangle \sim \lambda_{\rm env}\hbar^2/\tau_{\rm D}$,
 see Eq.~(\ref{dephi_rate}). This induces a typical force
 $\sim \langle {\cal U}^2\rangle^{1/2}/\xi
 \sim (\hbar/\xi)(\lambda_{\rm env}/\tau_{\rm D})^{1/2}$ on the particle. 
 To see if this noisy force significantly 
 modifies the paths in the vicinity of the encounter, we compare it with the 
 relative force of the chaotic system Hamiltonian on the particle at 
 the encounter. 
 Since the perpendicular extension of the encounter
 is $\delta r_\perp \sim (L\lambda_{\rm F})^{1/2}$, and the duration
 of the encounter is of order the Lyapunov time $\sim \lambda^{-1}$,
 the system force goes like 
 $m\lambda^2 \delta r_\perp\sim m\lambda^2 (L\lambda_{\rm F})^{1/2}$.
 Estimating $\lambda^{-1} \sim v_{\rm F}/L$ as is typical of chaotic billiards,
 the ratio of the noisy force to the system force becomes
 $[ \lambda_{\rm env} L\lambda_F /  (\xi^2 \lambda^2 \tau_D)]^{-1/2} $.  
 Thus one can ignore the modifications of the classical paths due to
 the coupling to the environment, as long as 
 \begin{eqnarray}
 \xi \gg   (\lambda_{\rm F}L)^{1/2} \times 
 \left[ \lambda_{\rm env}/\lambda \over \lambda \tau_D \right]^{1/2} .
 \end{eqnarray}
 We see that $\xi$ can easily be less than the typical
 encounter size $\delta r_\perp \sim (L\lambda_{\rm F})^{1/2}$ 
 (remember that $\lambda\tau_{\rm D} \gg 1$ is always assumed).
 Thus our method is not only applicable for $\xi$ up to the system size,
 where dephasing happens only in the loop. 
 It is also applicable for $\xi$ smaller than
 the encounter size, in which case the time $T_\xi(\eps)$ is negative,
 and dephasing occurs in part of the legs as well as the whole of the loop, 
 see Fig.~\ref{fig:negative-tau_xi}.

 Finally we caution the reader that the whole semiclassical method
 used in this article relies on the lead width being greater than the encounter
 size, this requires that $\lambda \tau_{\rm D} \ll (L/\lambda_{\rm F})^{1/2}$,
 thus we {\it cannot} access the regime $\xi \sim \lambda_{\rm F}$, which is dominated
 by stochastic diffraction at the leads.

 %~~~~~~~~~~~~~~~~~~~~~~~~FANO FACTOR ~~~~~~~~~~~~~~~~~~~~~~~~~~~%

 \subsection{Shot noise in the presence of an environment} 

 When the temperature of the electrons in the leads coupled to 
 a chaotic system is taken to zero, there is no thermal noise in the current
 through the device.  However there is still the intrinsically quantum noise
 which originates from the wave-like nature of the electrons.
 This zero-temperature noise is known as shot noise~\cite{Blanter-review}.
 In the absence of dephasing,
 shot noise has been well-studied using RMT \cite{Been97},
 quasi-classical field theory \cite{agam},
 and semiclassical methods \cite{wj2004,Brau06,wj2005-fano}.

 It is generally argued that the shot noise 
 is unaffected by the presence of an environment
 which causes dephasing but not heating of the electrons -- the regime
 of {\it phase-breaking} of Ref.~[\onlinecite{vanLan97}].
 This belief is founded on the fact that (i) 
 the dephasing-lead model gives a dephasing-independent 
 shot noise,
 (ii) kinetic equations -- in which interference effects
 are ignored -- give the same shot noise as 
 full quantum calculations.
 Here we show explicitly that, under the assumption that the system-environment
 coupling does not heat up the
 current-carrying electrons, indeed, the coupling to the 
 environment does not affect shot noise.

\begin{figure}
 \begin{center}
 \includegraphics[width=8cm]{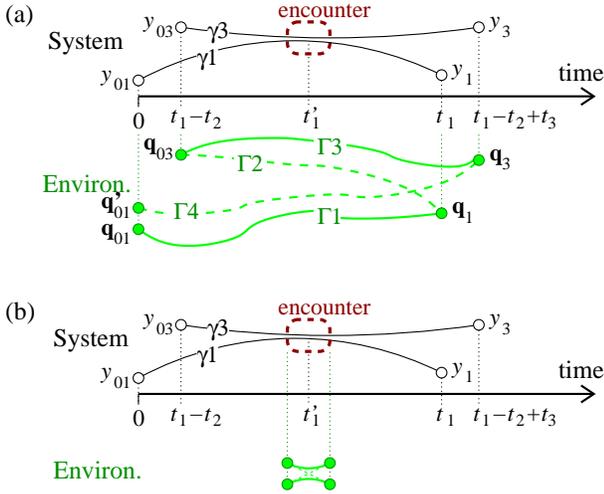}
 \caption{\label{fig:shot-noise}
 (color online) 
 Sketch of typical trajectories which contribute to shot noise
 in the presence of an environment.
 In both (a) and (b) the system (environment) 
 paths are sketched above (below) the time axis.
 In (a) we show a contribution which will survive system averaging, 
 because the system paths are paired almost everywhere
 with an encounter at time $t_1'$. There is  
 no constraint on the environment paths, as yet.
 For simplicity we show only system paths $\gamma1$ and $\gamma3$,
 with 
 path $\gamma2$ and $\gamma4$ being the same as $\gamma1$ and $\gamma3$ 
 except that they cross at the encounter. 
 Depending on the choice of $t_1'$ with respect to the other timescales,
 $t_1,t_2,t_3$, this pair of system paths
 could represent any of the 
 contributions in Fig.~1 of Ref.~[\onlinecite{wj2005-fano}].
 In (b) additional constraints are imposed on the $\Gamma_i$'s,
 after one has integrated over all
 initial and final positions of the environment.
 This integration removes all contributions from {\it environment}
 propagation when the {\it system} paths are paired. 
 }
 \end{center}
 \end{figure}

 We start with the formula for the zero-frequency 
 shot noise power through a system coupled to an environment.
 This formula is derived in  Appendix~\ref{appendix:current-noise},
 and is given as $S_{\rm RR}(0)$ in Eq.~(\ref{shot-noise-final}).
 We use Eq.~(\ref{sc-smatrix}) 
 to write each matrix element as sums over 
 classical paths. This gives us a sum over eight paths
 -- 4 system paths and 4 environment paths --
 as sketched in Fig.~\ref{fig:shot-noise}.
 The system paths are as follows
 \begin{eqnarray}
 & \bullet & \hbox{
 $\gamma1$ from $y_{01}$ on lead L to $y_1$ on lead R,}
 \nonumber \\
 & \bullet &\hbox{
 $\gamma2$ from $y_{03}$ on lead R to $y_1$ on lead R,}
 \nonumber \\
 & \bullet &\hbox{
 $\gamma3$ from $y_{03}$ on lead R to $y_3$ on lead R,} 
 \nonumber \\
 & \bullet & \hbox{
 $\gamma4$ from $y_{01}$ on lead L to $y_3$ on lead R.} \nonumber
 \end{eqnarray}
 The sums over lead modes and the trace over the environment density matrix
 are performed in the same manner as for the conductance, 
 [see above Eq.~(\ref{sc-cond})], which results in
 \begin{eqnarray}
 \label{eq:noise1}
 \hskip -4mm
 S \!
 &=& 
 {e^3V \over (2\pi \hbar)^3}
 \!\int_{\rm L} \! \! \rmd y_{01} \int_{\rm R} \! \rmd y_{03} \rmd y_1 \rmd y_3 
 \sum_{\gamma1,\cdots \gamma4} 
 A_{\rm sys}
 \e^{\rmi\Phi_{\rm sys}} 
 \nonumber \\
 \hskip -4mm
 & & \! \!
 \times
 \int \rmd {\bf q}_{01} \rmd {\bf q}_{03} \rmd {\bf q}_1 \rmd {\bf q}_3
 \! \!
 \sum_{\Gamma1,\cdots \Gamma4} \!
 A_{\rm env} 
 \e^{\rmi(\Phi_{\rm env}+\Phi_{\cal U})} .
 \end{eqnarray}
 Here, $A_{\rm sys} = A_{\gamma1}A_{\gamma2}A_{\gamma3}A_{\gamma4}$,
 $\Phi_{\rm sys} = (S_{\gamma1}-S_{\gamma2}+S_{\gamma3}-S_{\gamma4})/\hbar$
 and we absorbed all Maslov indices into the actions.
 Similarly, $A_{\rm env} = A_{\Gamma1}A_{\Gamma2}A_{\Gamma3}A_{\Gamma4}$,
 $\Phi_{\rm env} = (S_{\Gamma1}-S_{\Gamma2}+S_{\Gamma3}-S_{\Gamma4})/\hbar$,
 and $\Phi_{\cal U} =({\cal S}_{\gamma1,\Gamma1} -{\cal S}_{\gamma2,\Gamma2}
 +{\cal S}_{\gamma3,\Gamma3}  -{\cal S}_{\gamma4,\Gamma4})/\hbar$.
 As argued above, in the regime of pure dephasing,
 $\Phi_{\rm sys}$, $\Phi_{\rm env}$ and $\Phi_{\cal U}$ are uncorrelated.
 We thus first pair up the system paths to minimize
 $\Phi_{\rm sys}$, this pairing
 is the same as it would be in the absence of the environment
 (compare Fig.~\ref{fig:shot-noise} to Fig.~1 of 
 Ref.~[\onlinecite{wj2005-fano}]). 
 We see from the construction of  Eq.~(\ref{shot-noise-final}),
 that all paths reach the encounter 
 at the same time\cite{footnote:shot-noise-time-shift}, $t_1'$.

 %%%%%%%%%%%%%%%%%%%%%%%%%%%%%%%%%%%%%%%%% %%%%%%%
 %%%%%%%%%%%%%%%%%%%%%%%%%%%%%%%%%%%%%%%%% %%%%%%%

 Now we make the crucial observation that,
 for any given set of system paths, we have $\gamma1 \simeq\gamma2$
 for times greater than $t_1'$.
 Thus for times greater than $t_1'$ we can write the sum over
 $\Gamma1,\Gamma2$ in the second line
 of Eq.~(\ref{eq:noise1})
 as 
 \begin{eqnarray}
 & & \hskip -5mm
 \int \rmd {\bf q}_1 \sum_{\Gamma1,\Gamma2} A_{\Gamma1}A_{\Gamma2} 
 \, \e^{\rmi (S_{\Gamma1}-S_{\Gamma2} + {\cal S}_{\gamma1,\Gamma1}
 - {\cal S}_{\gamma2,\Gamma2}) /\hbar}
 \nonumber \\
 &\simeq& \int \rmd {\bf q}_1\sum_{\Gamma1,\Gamma2} A_{\Gamma1}A_{\Gamma2} 
 \, \e^{\rmi (S_{\Gamma1}-S_{\Gamma2} + {\cal S}_{\gamma1,\Gamma1}
 - {\cal S}_{\gamma1,\Gamma2}) /\hbar}
 \nonumber \\
 &=& \int \rmd {\bf q}_1\ 
 \big[{\cal K}_{\rm env}'({\bf q}_1,t_1;{\bf q}_1'',t_1')\big]^* 
 {\cal K}_{\rm env}' ({\bf q}_1,t_1;{\bf q}_1',t_1') \qquad
 \label{eq:Kprimes}
 \end{eqnarray}
 where we set $\gamma2=\gamma1$ to get the second line.
 We define ${\cal K}_{\rm env}' ({\bf q}_1,t_1;{\bf q}_1',t_1')$
 as the propagator for the environment evolving under an effective
 time-dependent potential, $V'(q,t)$,
 which is the sum of  ${\cal U}(r_{\gamma1}(t),q)$ and the potential term in $H_{\rm env}$.
 Since both propagators in Eq.~(\ref{eq:Kprimes}) evolve under the same 
 potential (because $\gamma1=\gamma2$), 
 we can use the basic property of propagators\cite{Feynman-Hibbs-book}
 that
 \begin{eqnarray}
 & & \int \rmd {\bf q}_1\  
 \big[{\cal K}_{\rm env}'({\bf q}_1,t_1;{\bf q}',t_1')\big]^* 
 {\cal K}_{\rm env}' ({\bf q}_1,t_1;{\bf q}'',t_1')
 \nonumber \\
 & & \qquad \qquad = \de({\bf q}''-{\bf q}') 
 \label{eq:pairing-propagators}
 \end{eqnarray}
 to integrate out these propagators for times greater than $t_1'$.
 In their place we have a constraint that path $\Gamma1$ and $\Gamma2$
 must be the same at time $t_1'$.
 This is sketched in Fig.~\ref{fig:shot-noise};
 the paths $\Gamma1$ and $\Gamma2$ 
 to the right of the encounter in Fig.~\ref{fig:shot-noise}a
 are replaced in Fig.~\ref{fig:shot-noise}b 
 by the constraint that
 paths $\Gamma1$ and $\Gamma2$ meet at time $t_1'$.
 Note that we cannot use Eq.~(\ref{eq:pairing-propagators}) 
 to integrate out paths $\Gamma1$ and $\Gamma2$ for arbitrary times before 
 $t_1'$ because the system paths $\gamma1$ and $\gamma2$ are then
 different enough that the potential $V'(q,t)$
 will be different for the two propagators in Eq.~(\ref{eq:Kprimes}).  
 However we can use the same argument to integrate out the pair 
 $\Gamma3$-$\Gamma4$ after time $t_1'$, and to integrate out the
 pairs $\Gamma1$-$\Gamma4$ and $\Gamma2$-$\Gamma3$
 {\it before} the time $t_1'$.  
 After all these pairs are replaced by $\de$-functions, we get
 the situation shown in Fig.~\ref{fig:shot-noise}b.

 Focusing on $\xi$ much greater than the encounter size,
 the above method can be used to integrate
 out the environment paths for all $t>t_1'$ and all $t<t_1'$.
 This leaves a single point (the environment state at $t=t_1'$) 
 to integrate over.  Doing this we see that Eq.~(\ref{eq:noise1})
 reduces to 
 \begin{eqnarray}
 \label{eq:noise-without-env}
 S
 =
 {e^3V \over (2\pi \hbar)^3}
 \!\int_{\rm L} \! \! \rmd y_{01} \int_{\rm R} \! \rmd y_{03} \rmd y_1 \rmd y_3 
 \sum_{\gamma1,\cdots \gamma4} 
 A_{\rm sys}
 \e^{\rmi\Phi_{\rm sys}}. \quad 
 \end{eqnarray}
 This is identical to the shot noise formula in the absence of an environment.
 Thus we have completely removed the environment from the problem
 without affecting the shot noise of the system at all.
 To calculate the shot noise now, one simply needs to follow
 the derivation for a system without an environment in 
 Ref.~[\onlinecite{wj2005-fano}].  The result is most conveniently written
 in  terms of the Fano-factor, $F$, 
 which is the ratio of the shot noise to the Poissonian noise
 $2e\langle I \rangle$,
 where $\langle I \rangle = 2e^2g^{\rm D}V/h$ is the average current.
 One gets
 \begin{eqnarray}
 F &\equiv& S/2e\langle I\rangle 
 = {N_{\rm L} N_{\rm R}\over (N_{\rm L}+ N_{\rm R})^2} 
 \exp[-\tau_{\rm E}^{\rm op}/\tau_{\rm D}].
 \label{eq:Fano}
 \end{eqnarray} 
 This is of course independent of the coupling to the environment.

 As a final comment, we note that above we kept only the
 leading ${\cal O}[N]$ term in the shot noise.
 There is a hierarchy of weak-localization-like 
 corrections ${\cal O}[N^a]$, $a=0,-1, \ldots$
 to this result\cite{Brau06}, which
 are suppressed by dephasing in much the same way
 as the weak-localization correction to conductance.
 Thus for $\tau_{\rm E}^{\rm op} \ll \tau_{\rm D}$,
 we can expect the environment to cause a cross-over from
 the result in Ref.~[\onlinecite{Brau06}] to the result in
 Eq.~(\ref{eq:Fano}) with $\tau_{\rm E}^{\rm op}=0$.
 Hence in the classical limit of wide leads ($N_{\rm L,R} \gg 1$) 
 the environment's effect is negligible,
 however for narrow leads ($N_{\rm L,R} \sim 1$) 
 the environment's effect may be significant.

 %~~~~~~~~~~~~~~~~~~~~~~~~ CLASSICAL NOISE ~~~~~~~~~~~~~~~~~~~~~~~~~~~%

 \section{Classical Noise}
 \label{sect:classical}

 We have shown that, to capture the effect of dephasing on weak localization,
 it is sufficient to treat the environment at the level of the 
 diagonal approximation. We thus observe
 that, in the semiclassical limit of short wavelength, 
 $\lambda_{\rm F}/L_{\rm env} \to 0$, a quantum chaotic environment 
 has the same dephasing effect on weak-localization
 as the equivalent {\it classical} chaotic environment.
 Because correlations typically decay exponentially fast in 
 classical hyperbolic systems, this makes the effect of this classical
 environment very similar to a
 classical noise field with a suitably chosen
 spatial and temporal correlation function.
 In this Section we show that the conclusions that we 
 draw for a quantum environment can also be drawn for 
 a classical noise field.
 One common experimental example of such a field 
 is microwave radiation, applied to the chaotic dot either by accident or
 on purpose \cite{Marcus-expt-microwaves}. 
 A second example, is the common theoretical treatment of 
 electron-electron interactions
 as a source of classical (Johnson-Nyquist) noise \cite{Alt82,Cha86}.

 We add a new term to the 
 system Hamiltonian of the form, $V_{\rm noise}(t)$.
 We assume this term is weak enough that it does not affect the classical paths,
 but strong enough to modify the phase acquired along such paths.
 The phase difference for a pair of paths contributing to the conductance
 is $(\Phi_{\rm sys} +\Phi_{\rm noise})$,
 where $\Phi_{\rm sys}$ is given in Eq.~(\ref{PhiS})
 and 
 \begin{eqnarray}
 \Phi_{\rm noise} 
 = \int \rmd t \big[ V({\bf r}_{\gamma'};t) - V({\bf r}_{\gamma};t) \big]/\hbar.
 \end{eqnarray}
 We now assume that the noise is Gaussian distributed with
 \begin{eqnarray}\label{eq:classical-noise-correlation}
 & & \hskip -5mm
 \big\langle V\big({\bf r}_{\gamma'}(t);t\big) 
 V\big({\bf r}_{\gamma}(t');t'\big)\big\rangle 
 \nonumber \\
 &=& 
 \langle V^2\rangle \
 K_{\rm noise}\!\left({|{\bf r}_2-{\bf r}_1|/\xi}\right)\, 
 J_{\rm noise}(\lambda_{\rm noise}|t_2-t_1|) . \qquad
 \end{eqnarray}
 Here, $K_{\rm noise}(x)$ gives the form of the spatial decay of 
 the correlation function (on a scale $\xi$), 
 and  $J_{\rm noise}(t)$ gives the form of the temporal decay of 
 the correlation function (on a scale which we call $\lambda_{\rm noise}^{-1}$
 to make the analogy with the notation in Section ~\ref{sect:weak-loc-env}).
 We can now follow the derivation in Section ~\ref{sect:weak-loc-env})
 by replacing ${\cal U}({\bf r}_\gamma(t),{\bf q}_\Gamma(t))$ with 
 $V({\bf r}_\gamma;t)$ throughout.
 We assume
 that the correlations in time are short enough to be treated 
 as white-noise-like,
 $J_{\rm noise}(x)\propto\de(x)$, and that the spatial correlations 
 decay fast enough that we can justify the
 approximation in Eq.~(\ref{eq:f-approx}).
 This directly leads to 
 the same result for weak-localization as in Eq.~(\ref{eq:gwl-env}),
 where now
 \begin{eqnarray}
 \tau_{\phi}^{-1} \sim
 \hbar^{-2} \lambda_{\rm noise}^{-1}
 \langle V^2 \rangle.
 \end{eqnarray}

 \subsection{Noise due to electron-electron interactions
 in a 2-dimensional ballistic system.}
 \label{sect:e-e-2D}

 Here we consider the noise generated by  electron-electron interactions 
 in a ballistic chaotic system.
 In such a system, dephasing is caused by 
 noise with momenta ($\delta p$-vectors) 
 larger than the inverse system size, $L^{-1}$.
 Thus the dephasing processes are the same as those for 
 ballistic motion  in disordered systems ($\delta p$-vectors greater than 
 inverse mean free path). 
 Such processes were first studied in Ref.~[\onlinecite{Fukuyama83}]
 for spinless electrons, 
 while more recently  Ref.~[\onlinecite{Narozhny02}] explored the full crossover from
 ballistic to diffusive motion for electrons with spin.
 The effect of a finite Ehrenfest time on such dephasing 
 was considered in Ref.~[\onlinecite{Bro06-quasi-ucf}], which found 
 that the dephasing rate in the vicinity of the encounter 
 has a logarithmic dependence on the perpendicular distance between paths.
 This led them to observe that the electron-electron interaction in
 a 2-dimensional ballistic system 
 dephases weak-localization exponentially
 with the Ehrenfest time.
 We repeat their derivation here and show that 
 \begin{eqnarray}
 g^{\rm wl}_{\rm e-e} = 
 \frac{g^{\rm wl}_0 }{1+\tau_{\rm D}/\tau_\phi}
  \; \exp\left[-\left(\tEc+\half \tau_{L_{\rm T}}\right) /\tau_\phi\right],
 \label{eq:dephasing-e-e-interaction}
 \end{eqnarray}
 where
 $\tau_{L_{\rm T}}$ is given by Eq.~(\ref{eq:gwl-env}) with 
 $\xi$ equaling a thermal length scale 
 $L_{\rm T} = \hbar v_{\rm F}/k_{\rm B} T$.
 Ref.~[\onlinecite{Bro06-quasi-ucf}] neglected the $\tau_{L_{\rm T}}$-term 
 in the exponent since it is often small.

 In our qualitative derivation of this result,
 we treat the electron-electron interaction as classical-noise, 
 rather than using the perturbative field theory approach in 
 Refs.~[\onlinecite{Fukuyama83,Narozhny02,Bro06-quasi-ucf}]. 
 Our approach is similar in spirit to those for diffusive 
 systems~\cite{Alt82,Cha86}.
 The screened electron-electron interaction gives a correlation function 
 of the form \cite{Fukuyama83}
 \begin{eqnarray}
 \langle V_{\delta p,\omega}^2 \rangle 
 \propto {\delta \big( \omega-m^{-1}{\bf p}\cdot{\bf \delta p}\big) 
 \over |{\bf \delta p}|}
 \times J(\om),
 \label{eq:V-e-e1}
 \end{eqnarray}
 where we assume $|{\bf \delta p}| \ll |{\bf p}|$, so the energy difference
 between a particle with momentum $({\bf p}+{\bf \delta p})$ and ${\bf p}$ is
 ${\bf p}\cdot{\bf \delta p}/m$.
 The factor of $1/|{\bf \delta p}|$ comes from the imaginary part of the
 screened Coulomb interaction,
 is due to the polarization bubble
 and corresponds to the fluctuations of the electron sea at 
 momentum and energy $(\delta p,\omega)$.
 The $\delta$-function ensures that energy and momentum are conserved
 in the interaction between the system and the environment.
 The function $J(\om)$ gives the weight of environment modes 
 excited at energy $\om$. At temperature $T$ 
 it is typically of the form
 $J(\om) \approx \big[\sinh (\om/k_{\rm B} T)\big]^{-1}$.
 It is convenient to write ${\bf \delta p}$ in terms of components parallel and 
 perpendicular to the relevant classical path, i.e.~parallel/perpendicular to
 ${\bf p}$.
 Then,
 \begin{eqnarray}
 \langle V_{\delta p,\omega}^2 \rangle 
 \propto {\delta \big( \omega-v_{\rm F}\delta p_\parallel \big) \over
 [\delta p_\parallel^2 + \delta p_\perp^2]^{1/2}}
 \times J(\om),
 \label{eq:V-e-e}
 \end{eqnarray}
 and we have 
 \begin{subequations}\label{eq:noise-correlator-d}
 \begin{eqnarray}
 & & \hskip -5mm
 \langle V({\bf r}_{\gamma'}(t');t') V({\bf r}_{\gamma}(t);t)\rangle 
 \nonumber \\
 &=& 
 \int \rmd^d {\bf \delta p} \int \rmd \om
 \e^{\rmi [{\bf \delta p}\de{\bf r}(t',t) 
 + \om (t'-t)]/\hbar}\
 \langle V^2_{{\bf \delta p},\om} \rangle 
 \\
 &\propto& \int \rmd \om J(\om) \exp \big[ \rmi 2\om (t'-t)/\hbar \big] 
 \nonumber \\
 & & \times {\rm Re} \left[\int_{\hbar/L}^{p_{\rm F}}  \rmd \delta p_\perp 
 {\exp [\rmi \delta p_\perp \delta r_\perp/\hbar ]
 \over [\delta p_\perp^2 +(\om/v_{\rm F})^2]^{1/2}} \right].
 \label{eq:q-perp-int}
 \end{eqnarray} 
 \end{subequations}
 Here, $\de{\bf r}(t',t)=
 ({\bf r}_{\gamma'}(t')-{\bf r}_{\gamma}(t))$.
 To get the second line, we wrote 
 $\de {\bf r}(t',t) = (\delta r_\parallel,\delta r_\perp)$ 
 in the basis parallel and perpendicular to ${\bf p}$,
 we then inserted Eq.~(\ref{eq:V-e-e}),
 evaluated the $\delta p_\parallel$-integral and noted that
 $\delta r_\parallel= v_{\rm F}(t'-t)$.
 Eq.~(\ref{eq:noise-correlator-d}) expresses the correlator of the
 Coulomb interaction along classical trajectories in the form convenient
 for our semiclassical approach, Eq.~(\ref{eq:classical-noise-correlation}).
 At first sight the form of the interaction 
 in Eqs.~(\ref{eq:V-e-e1},\ref{eq:V-e-e}) does not appear to correspond
 to a classical noise-field, since it is a function of the
 momentum, ${\bf p}_\gamma(t)$, of {\it system} paths.
 The correlator must however be evaluated on weak localization
 loops, in which case one can use the fact that
 $r_{\parallel \gamma'}(t)= r_{\parallel \gamma}(t)
 = v_{\rm F}t$ throughout the encounter region to perform
 the Fourier transform.
 Thus 
 the interaction term in Eq.~(\ref{eq:q-perp-int})
 is equivalent to a classical-noise field which is a single function 
 of $\delta {\bf r}$ and $t$ for all ${\bf p}_\gamma$.
 This function must simply be chosen such that the integral over
 $\delta p_\pll$ reduces to  Eq.~(\ref{eq:q-perp-int}).

 The time-dependent part of the correlator
 is given by the $\omega$-integral in
 Eq.~(\ref{eq:q-perp-int}). We assume that the 
 temperature is high enough, $k_{\rm B} T > \hbar v_{\rm F}/L$, 
 that the correlation time becomes shorter than the time of flight
 $L/v _{\rm F}$ through the cavity. Accordingly, we treat the noise 
 as $\delta$-correlated in time, and
 set $t'=t$ from now on.

 We next investigate the properties of the real part 
 of the $\delta p_\perp$-integral in Eq.~(\ref{eq:q-perp-int}),
 giving the spatial dependence of the correlator.
 We write it as
 \begin{eqnarray}
 G(\de r_\perp) 
 = {\rm Re} \left[
 \int_{\de r_\perp/L}^{p_{\rm F}\de r_\perp/\hbar}
 {\rmd x \, \e^{\rmi x} \over [x^2 +(\de r_\perp/L_\om)^2]^{1/2}} \right]
 \, ,
 \end{eqnarray}
 where $L_\om = \hbar v_{\rm F}/\om$ is the distance a system
 particle will travel on the timescale that the $\om$-energy
 component of the noise fluctuates.
 For the ballistic model of the e-e interactions to be valid we need
 that $L_\om \ll L$, but we assume that $L_\om \gg \lambda_{\rm F}$.
 We can easily evaluate this integral in the following regimes
 \cite{footnote:integrals},
 \begin{eqnarray} 
 \label{eq:G}
 G(\delta r_\perp)\simeq 
 \left\{ \begin{array}{ccc}
 \ln [L_\om / \lambda_{\rm F}], & \quad & 
 \hbox{for } \de r_\perp \ll \lambda_{\rm F},  \\
 \ln [L_\om / \de r_\perp], & \quad & 
 \hbox{for } \lambda_{\rm F} \ll \de r_\perp \ll L_\om, 
 \\
 0,  & \quad & 
 \hbox{for }  L_\om \ll \de r_\perp \sim  L,
 \end{array}\right.
 \end{eqnarray}
 where we neglected all ${\cal O}[1]$-terms
 (for example the result for $\de r_\perp \gg  L_\om$ is actually 
 ${\cal O}[L_\om/\de r_\perp] \lesssim {\cal O}[1]$).
 The crossover between these regimes is smooth.
 We thus conclude that $K_{\rm noise}(x)$, as defined by 
 Eq.~(\ref{eq:classical-noise-correlation}), becomes
 \begin{eqnarray}
 K_{\rm noise}(\de r/L_\om) 
 = {\ln[L_\om/\de r_\perp] \over \ln [L_\om/\lambda_{\rm F}]}
 \end{eqnarray}
 This function does {\it not} decay fast as $\de r_\perp$ grows, 
 and thus it {\it cannot} be treated
 in the manner we do elsewhere in this article, i.e. we cannot write
 an equation analogous to Eq.~(\ref{eq:f-approx}). 
 Instead we note that the dephasing rate in the vicinity of the encounter
 (where $\lambda_{\rm F}  \ll \de r_\perp \ll L_\om$)
 goes like 
 $\big[G(0)-G(\lambda_{\rm F}  \ll\de r_\perp \ll L_\om)\big]
 = \ln [\de r_\perp/\lambda_{\rm F}]$.
 While the dephasing rate in the loop, 
 $\tau_\phi^{-1}$, goes like 
 $\big[G(0)-G(\de r_\perp \sim L_\om)\big]
 \sim \ln [L_\om/\lambda_{\rm F}]
 $.
 Now we note that the integral over $\om$ is dominated by
 $\om \simeq k_{\rm B} T \ll E_{\rm F}$, 
 thus we can define
 $L_{\rm T} = \hbar v_{\rm F}/k_{\rm B} T$ and
 write
 $\ln [L_\om/\lambda_{\rm F}] = \ln [L_{\rm T}/\lambda_{\rm F}] 
 - \ln [\om/k_{\rm B} T]$. 
 We can neglect the second term as it is much smaller than the first, and
 thereby replace $L_\om$ with $L_{\rm T}$ in the above formulae. 
 In this way we
 reproduce the result in Ref.~[\onlinecite{Bro06-quasi-ucf}], 
 that the dephasing rate in the vicinity of the encounter 
 is
 \begin{eqnarray}
 \tilde{\tau}_\phi^{-1}(\de r_\perp \ll L_{\rm T}) 
 = \tau_\phi^{-1} {\ln [\de r_\perp/\lambda_{\rm F}] 
 \over \ln [L_{\rm T}/\lambda_{\rm F}] }
 \end{eqnarray}
 while 
 $\tilde{\tau}_\phi^{-1}(\de r_\perp \gtrsim L_{\rm T}) = \tau_\phi^{-1}$.
 This is rather different from 
 the systems considered elsewhere in this article
 where the dephasing rate is approximately zero  for $\de r_\perp < \xi$ and 
 approximately constant for $\de r_\perp > \xi$.

 We now calculate the effect of such a $\de r_\perp$-dependent
 dephasing rate.
 We note that close to the encounter, 
 $\de r_\perp = \half \eps L \e^{\lambda \tau}$ 
 where $\tau$ is the time measured
 from the encounter.  
 We split the dephasing into two contributions. 
 The first contribution is where
 $\delta r_\perp \geq L_{\rm T}$ 
 (here dephasing is time-independent, at the rate $\tau_\phi^{-1}$),
 the second is where the paths have 
 $\lambda_{\rm F} < \delta r_\perp < L_{\rm T}$
 (here dephasing is time-dependent).
 The boundary between the two contributions is at $\tau=T_{\rm T}(\eps)/2$,
 where we define $T_{\rm T}(\eps)= \lambda^{-1}\ln [L_{\rm T}^2/(L\eps)^2]$.
 The lower bound on the second contribution ($\delta r_\perp=\lambda_{\rm F}$)
 is at time $\tau= T_{\rm T}(\eps)/2-\lambda^{-1}\ln[L_{\rm T}/\lambda_{\rm F}]$. 
 Then the exponent induced by the dephasing is  
 \begin{eqnarray}
 & & \hskip -5mm 
 -{t_2-t_1 - T_{\rm T}(\eps) \over \tau_\phi} 
 -2\int_{T_{\rm T}/2-\lambda^{-1}\ln[L_{\rm T}/\lambda_{\rm F}]}^{T_{\rm T}/2} 
 {\rmd \tau \over \tilde{\tau}_\phi
 \big(\de r_\perp \big)}
 \nonumber \\
 &=& \! - {t_2-t_1 - T_{\rm T}(\eps) \over \tau_\phi }
 -{\lambda^{-1} \ln[L_{\rm T}/\lambda_{\rm F}]\over \tau_\phi} \qquad
 \end{eqnarray}
 where to evaluate the integral we defined $\tau'=\tau-T_{\rm T}(\eps)/2$.
 The first term (which comes from the dephasing in the loop)
 alone would give no exponential term in the dephasing.
 The integral of that term over $t_2-t_1$ gives it a form
 $T_{\rm L}(\eps)-T_{\rm T}(\eps) = \tau_{L_{\rm T}}$
 giving an exponent like in Eq.~(\ref{eq:gwl-env}) with $\xi=L_{\rm T}$.
 However we also have the second term which gives dephasing 
 in the vicinity of the encounter, 
 we can write it in terms of an Ehrenfest time using 
 $\lambda^{-1}\ln[L_{\rm T}/\lambda_{\rm F}] = \tau_{\rm E}^{\rm cl} 
 -\half \tau_{L_{\rm T}}$.
 Summing the two terms we find that the exponential term in the dephasing
 goes like $\tilde{\tau}/\tau_\phi$,
 with  $\tilde{\tau} = \tau_{\rm E}^{\rm cl} 
 +\half \tau_{L_{\rm T}}$. 
 This is the result 
 which we gave in Eq.~(\ref{eq:dephasing-e-e-interaction})
 and was found in Ref.~[\onlinecite{Bro06-quasi-ucf}]
 (neglecting the $\tau_{L_{\rm T}}$-term).

 Because $\lambda_{\rm F}$ is the scale of Friedel oscillations, 
 one might have expected that electron-electron interactions
 give a noise with a correlation length $\xi \sim \lambda_{\rm F}$,
 which would lead to a suppression $\propto \exp[-2\tEc/\tau_\phi]$ 
 of weak localization, instead of $\exp[-\tEc/\tau_\phi]$.   
 The factor of 2 difference
 between the correct result and this naive argument,
 is due to the fact that 
 all scales (i.e. all values of $\delta p$) contribute to the
 noise induced by the electron-electron interactions. 

 \subsection{Noise due to the coupling to the electrostatic environment}

 %%%%%%%%%%%%%%%%%%%%%%%%%%%%%%%%%%%%%%%%% %%%%%%%
 \begin{figure}
 \begin{center}
 \includegraphics[width=8cm]{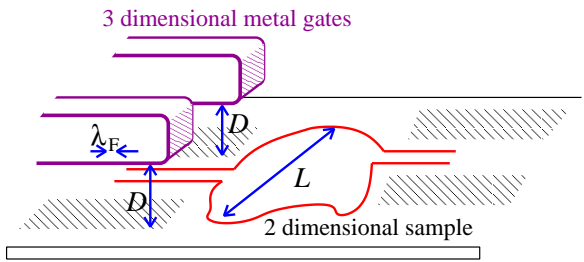}
 \caption{\label{fig:3d-gates} 
 Sketch of a typical chaotic quantum dot in a 
 two-dimensional electron gas (2deg).
 The dot is defined by metallic gates which 
 are biased to deplete the 2deg everywhere except in the 
 regions defining the leads and the dot.
 These gates are a distance $D$ above the 2deg 
 (with $D \gg \lambda_{\rm F}^{\rm gate}$). 
 At finite temperature, the electrons at the surface of these
 metallic gates will fluctuate, leading to noise which will be felt by
 the electrons in the chaotic quantum dot.
 }
 \end{center}
 \end{figure}
 %%%%%%%%%%%%%%%%%%%%%%%%%%%%%%%%%%%%%%%%% %%%%%%%

 Our aim here is to show that electron-electron interactions do {\it not} 
 automatically lead to dephasing which is an exponential function of the
 Ehrenfest time.  It only happens when the $q$-integral is divergent at its
 upper limit, with an 
 upper cut-off of order $p_{\rm F}$. 
 If the $q$-integral is cut-off by some other length scale,
 then the dephasing of weak-localization will be independent of the 
 Ehrenfest time.

 The example we consider is noise in a two-dimensional system
 due to electron-electron interactions between the system and the gates. 
 A typical experimental set-up is sketched in Fig.~\ref{fig:3d-gates}.
 The gates are bulk metal, the electrons
 in the system will feel fluctuations of electrons at the surface of the
 gates.  One can expect that these fluctuations
 at the surface of the gate are sufficiently well-confined 
 to two-dimensions (by screening in the bulk of the gate),
 that they will cause a noise field in the system of a form similar to
 Eq.~(\ref{eq:V-e-e1}). 
 However the fact that the distance between the gates and the chaotic system
 is $D$ means that the natural upper cut-off
 on the $q$-integral will be $\hbar/D$ and not $p_{\rm F}$.
 Assuming $D \ll L,L_\om$, we replace $p_{\rm F}$ by $\hbar/D$ 
 throughout the derivation in Section~\ref{sect:e-e-2D}, 
 and find that
 \begin{eqnarray}
 \label{eq:gwl-gate}
 g^{\rm wl}_{\rm gate \, e-e} = 
 \frac{g^{\rm wl}_0 }{1+\tau_{\rm D}/\tau_\phi}
  \; \exp[-\tau_\xi/\tau_\phi], 
 \end{eqnarray}
 with $\xi=(L_{\rm T}D)^{1/2}$ and hence 
 $\tau_\xi = \lambda^{-1}\ln[L^2/(L_{\rm T}D)]$.
 The dephasing rate here is $\tau_\phi^{-1} \propto D^{-1}$.
 In systems in which dephasing is dominated by
 thermal system-gate interactions, 
 we therefore expect a dephasing that is $\tE$-independent and
 algebraic in $\tau_\phi$ for $D\sim L$. 

 In real experiments the gates are typically much more disordered than
 the chaotic system, thus we can easily have a situation in which
 the thermally excited modes in the gates are diffusive -- or even localized
 by a charge trap at the edge of a gate -- in which case
 their noise field will be similar to that in 
 Ref.~[\onlinecite{Alt82,Cha86}].  This modifies the integrand of the
 $q$-integral, however the upper cut-off will still be $\hbar/D$, 
 therefore dephasing will again be $\tE$-independent.

 In general, dephasing is due to a combination
 of e-e interactions {\it within} the system 
 and the Coulomb coupling between the system and external charge 
 distributions in gates and other reservoirs of charges. Which of these sources
 of dephasing dominates is determined by microscopic details which
 we do not discuss here, 
 in particular the temperatures and mean-free-paths of both system and 
 gates~\cite{Fukuyama83}.

  %~~~~~~~~~~~~~~~~~~~~~~~~DEPHASING LEAD MODEL ~~~~~~~~~~~~~~~~~~~~~~~~~~~%
 \section{DEPHASING LEAD MODEL }\label{DEPHAS-SEC}

\begin{figure}
 \begin{center}
 \includegraphics[width=4.2cm]{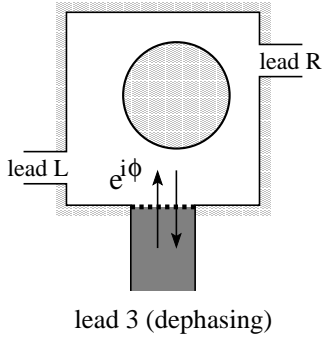}
 \caption{\label{fig:lead_model} 
 Schematic of the dephasing-lead model.
 The system is a open quantum dot with an extra lead (lead 3), 
 whose voltage is chosen
 to make zero current flow (on average) in that lead.
 This extra lead thereby 
 causes dephasing without loss of particles.
 }
 \end{center}
 \end{figure}

 In its simplest formulation the dephasing lead model consists of 
 adding a fictitious lead $3$ to the cavity.
 This is illustrated in Fig.~\ref{fig:lead_model}.  
 Contrary to  the two real leads L,R, the potential  voltage on lead $3$  
 is tuned
  such that the net current through it is zero. Every electron that leaves 
 through lead $3$ 
 is replaced by one with an unrelated phase, leading to a loss of phase 
 information without loss of current.

 In this situation the conductance from L to R is given by~\cite{But86} 
 \begin{equation}\label{3leadgeo}
 g = T_{\rm RL} + \frac{T_{\rm R3}\,T_{\rm 3 L}}{T_{\rm 3L}+T_{\rm 3R }}, 
 \end{equation}
 where $T_{nm}$ is the conductance from lead 
 $m$ to lead $n$  in the absence of a voltage on lead $3$.
 We separate the Drude and weak-localization parts
 of $T_{nm}$, 
 \begin{equation}\label{splitdrude-wl}
 T_{nm}=  T_{nm}^{\rm D}+ \delta T_{nm}+ {\cal O}[N_{\rm T}^{-1}],\,
 \end{equation}
 where the Drude  contribution,  $T_{nm}^{\rm D}$, is ${\cal O}[N_{\rm T}]$ 
 and the weak-localization contribution, $\delta T_{nm}$, 
 is ${\cal O}[N_{\rm T}^0]$ and 
 $N_{\rm T} =N_{\rm L} + N_{\rm R} +N_{3}$ is the total number of channels 
 in this three terminal geometry.
 We expand $g$ for large $N_{\rm T}$ and collect all  
 ${\cal O}[N_{\rm T}]$-terms
 (Drude contributions) and all  ${\cal O}[N_{\rm T}^0]$-terms
 (weak-localization contributions) to get $g= g^{\rm D}+g^{\rm wl}$ with 
 \begin{subequations}
 \begin{eqnarray}
 g^{\rm D} &=& T^{\rm D}_{\rm RL} + 
 \frac{T^{\rm D} _{\rm R3}T^{\rm D} _{\rm 3L} } { T^{\rm D} _{\rm 3L}+T^{\rm D} _{\rm 3R}} \; ,
 \label{3leadmodelDr}\\
 g^{\rm wl}  
 &=& \delta T _{\rm RL}
  +  
 \frac{ 
 (T^{\rm D}_{\rm R3})^2 \delta T_{\rm 3L}
 +(T^{\rm D}_{\rm 3L})^2 \delta T_{\rm 3R} }
 { (T^{\rm D} _{\rm 3R} +T^{\rm D} _{\rm 3L})^2} \, .
 \label{3leadmodelWL}
 \end{eqnarray}
 \end{subequations}
 These equations form the basis of our semiclassical derivation of 
 weak localization in the dephasing lead model. We first consider the 
 case of a dephasing lead perfectly coupled to the cavity,
 and then move on to consider
 a dephasing lead with a tunnel barrier of transparency $\rho < 1$.
 We finally discuss multiple dephasing leads.

  %%%%%%%%%%%%%%%%%%%%%%%%%%%%%%%%%%%%%%%%% %%%%%%%
 %%%%%%%%%%%%%%%%%%%%%%%%%%%%%%%%%%%%%%%%% %%%%%%%

 %--------------------------------------------------------------------------------------------%
 \subsection{Dephasing lead without tunnel-barrier}

 \subsubsection{Weak-localization}

 The Drude conductance and weak-localization correction 
 from lead $m$ to lead $n$ in a three-lead cavity are
 \begin{eqnarray} 
 T_{nm}^{\rm D} &=&  \frac{N_{n}N_{m}}{N_{\rm T}},
 \\
 \delta T_{nm} &=& - \frac{ N_{n}N_{m}}{N_{\rm T}^2}
  \exp[-\tEc/\tilde{\tau}_{\rm D}]
 \end{eqnarray}
 where $\tilde{\tau}_{\rm D}^{-1} = (\tau_0L)^{-1} (W_{\rm L} + W_{\rm R}+W_3)$,
 in terms of $\tau_0$, the time of flight across the system.
 We substitute these results
 into Eq.~(\ref{3leadmodelDr}) and Eq.~(\ref{3leadmodelWL}), 
 and write the answer in terms 
 of the dwell time in the two lead (L and R) geometry, $\tau_{\rm D}$,
 and the dephasing rate, $\tau_\phi^{-1}$, which we define 
 as the decay rate to lead $3$,
 \begin{eqnarray}\label{tdandtphi}
 \tau_{\rm D}^{-1} &=& (\tau_0L)^{-1} (W_{\rm L}+W_{\rm R}),
 \\
 \tau_{\phi}^{-1} &=& (\tau_0L)^{-1} W_3.
 \end{eqnarray}
 We have ${\tilde\tau}_{\rm D}^{-1}= \tau_{\rm D}^{-1} + \tau_\phi^{-1}$, and 
 from this we find the Drude conductance and weak-localization 
 correction, 
    \begin{eqnarray}
    g^{\rm D} &=& g^{\rm D}_0, \\
 \label{eq:weakloc_3lead}
    g^{\rm wl} &=& \frac{g^{\rm wl}_0} { 1+ \tau_{\rm D}/\tau_{\phi} } \exp[-\tEc/\tau_{\phi}].
    \end{eqnarray}
 Here $g^{\rm D}_0$ and $g^{\rm wl}_0$ are the results for a two lead cavity 
 in the absence of dephasing, Eqs.~(\ref{eq:Drudecond})
 and (\ref{eq:gwl0})

 The weak-localization correction with a dephasing lead
 has a similar structure to that with a real environment.
 However here the time scale involved in the additional exponential suppression 
 contains no independent parameter analogous to $\xi$.
 We could have expected that 
 the width of the dephasing lead would play a role similar to $\xi$.
 However this turns out not to be the case, instead the Fermi wavelength appears
 in place of $\xi$, so the time scale in the additional exponential suppression
 is the Ehrenfest time, $\tEc$.

 \subsubsection{Universal conductance fluctuations}

 In the absence of tunnel barrier, we can go further and calculate
 conductance fluctuations at almost no extra cost.
 From Eqs.~(\ref{3leadgeo}) and (\ref{splitdrude-wl}), 
 we get the following expression for 
 the variance of the conductance to order ${\cal O}[N_{\rm T}^0]$
 \begin{eqnarray}
 {\rm var} \, g &=& {\rm var} \,T _{\rm RL} \nonumber \\
 &+&\frac{(T^{\rm D}_{\rm 3R})^4 \, {\rm var} \, T _{\rm 3L}  +  (T^{\rm D}_{\rm 3L})^4 \, 
 {\rm var} \, T _{\rm 3R}}{(T^{\rm D} _{\rm 3R} +T^{\rm D} _{\rm 3L})^4}
 \nonumber \\
 &+& 2 \,
  \frac{(T^{\rm D}_{\rm 3R} \, T^{\rm D}_{\rm 3L})^2 \,
 {\rm covar} \, (T _{\rm 3L}, T _{\rm 3R})}{(T^{\rm D} _{\rm 3R} +T^{\rm D} _{\rm 3L})^4}
 \nonumber \\
 &+& 
 2 \, \frac{ (T^{\rm D}_{\rm 3L})^2 \, {\rm covar}(T _{\rm RL},T_{\rm 3R})}
 {(T^{\rm D} _{\rm 3R} +T^{\rm
     D} _{\rm 3L})^2}
 \nonumber \\ 
 &+& 2 \, \frac{(T^{\rm D}_{\rm 3R})^2 \, {\rm covar}(T _{\rm RL},T_{\rm 3L}) 
 }{(T^{\rm D} _{\rm 3R} +T^{\rm
     D} _{\rm 3L})^2}.
 \end{eqnarray}
 In the universal regime, Ref.~[\onlinecite{Bar95}] gives, 
 to order ${\cal O}[N_{\rm T}^0]$ in the presence of time-reversal symmetry,
 \begin{eqnarray}
 {\rm var} T_{ij} &=& \frac{N_i^2 \, N_j^2}{N_{\rm T}^4}, \label{ucf:baranger}\\
 {\rm covar} \, (T_{ij},T_{ik}) &=& \frac{N_i^2 \, N_j \, N_k}{N_{\rm T}^4}.\label{covar:baranger}
 \end{eqnarray}
 Ref.[\onlinecite{Bro06-ucf}] showed that Eq.~(\ref{ucf:baranger})
 remains valid even at finite $\tE/\tD$. Inspection of their
 calculation for ${\rm var} T_{ij}$ shows that the same conclusion also applies
 to ${\rm covar} \, (T_{ij},T_{ik})$, and thus Eq.~(\ref{covar:baranger})
 still holds independently of $\tE/\tD$.
 Together with Eq.~(\ref{tdandtphi}), this straightforwardly leads to 
 \begin{eqnarray}\label{varg_double_lorentzian}
 {\rm var} \, g &=& \frac{N_{\rm R}^2 \, N_{\rm L}^2}{(N_{\rm R}+N_{\rm L})^4}
 \frac{1}{(1 +\tau_{\rm D}/\tau_{\phi})^2}.
 \end{eqnarray}
 In the universal regime, 
 this result was previously derived in Ref.~[\onlinecite{Brou97}].
 We thus conclude that, in the dephasing lead
 model without tunnel-barrier, conductance fluctuations exhibit the universal
 behavior of Eq.~(\ref{varg_double_lorentzian}). Below, we confirm this 
 result numerically.

 %--------------------------------------------------------------------------------------------%
 \subsection{Dephasing lead with tunnel-barrier}

 Putting a tunnel-barrier on the dephasing lead $3$ 
 is attractive because one can avoid the local character of the 
 dephasing lead model by considering a wide 
 third lead with an almost opaque barrier~\cite{Brou97}. Additionally,
 this is the model studied numerically in
 Ref.~[\onlinecite{Two04.2}] in the context of conductance fluctuations. 
 Weak-localization and shot noise in this model
 have been considered within the trajectory-based semiclassical
 approach in Ref.~[\onlinecite{Whi07}], and here we only mention
 the main results. 

 According to Ref.~[\onlinecite{Whi07}],  
 when all leads are connected to the cavity via tunnel-barriers,
 with the barrier on lead $m$ having transparency $\rho_m \in [0,1]$,
 the Drude conductance between lead
 $m$ and $n$, $T^{\rm D}_{nm}$, and
 the weak-localization correction, $\delta T_{nm}$, are
 \begin{eqnarray}
 T^{\rm D}_{nm}
 &=& \rho_n\rho_m N_nN_m/{\cal N},
 \\
 \delta T_{nm} 
 &=& \rho_n\rho_m \frac{N_nN_m}{{\mathcal N}^2}  
 \left(\rho_n+ \rho_m  - \frac{\tilde{\mathcal N }}{{\mathcal N}}\right)  
 \nonumber \\
 & & 
 \times \exp \left[-\tau_{\rm E}^{\rm op }/\tau_{\rm D_2} - 
 (\tau_{\rm E}^{\rm cl } -\tau_{\rm E}^{\rm op } )/\tau_{\rm D_1} \right]. 
 \quad \label{gijwl}
 \end{eqnarray}
 Here, $\tau_{\rm D1}^{-1} = (\tau_0L)^{-1} \sum_n \rho_n W_n$
 and $\tau_{\rm D2}^{-1} = (\tau_0L)^{-1} \sum_n \rho_n (2-\rho_n)W_n$
 are the single path and the paired paths survival times respectively,
 $W_n$ is the width of lead $n$, ${\mathcal N }= \sum_k \rho_k N_k$
 and $\tilde{\mathcal N }= \sum_k \rho_k^2N_k$.

 Now we assume that $\rho_{\rm L}=\rho_{\rm R}=1$ so only the dephasing lead has 
 a tunnel barrier. Substituting the Drude and weak-localization 
 contributions into Eq.~(\ref{3leadmodelWL}) we find that
 \begin{eqnarray}\label{eq:gwl3lead}
 g^{\rm wl} &=&  
 \frac{g^{\rm wl}_0}{ 1+\tau_{\rm D1} /\tau_{\phi}} 
 \exp\left[ -(1- \rho)  \tau_{\rm E}^{\rm op}/\tau_{\phi} -
 \tau_{\rm E}^{\rm cl}/\tau_{\phi} \right]. \qquad 
 \end{eqnarray} 
 The argument in the exponent in Eq.~(\ref{eq:gwl3lead})
 has a simple physical meaning.
 It is the probability that a path survives throughout the 
 paired-region ($\tEo/2$ on either side of the encounter)
 without escaping into lead $3$, 
 multiplied by the probability to survive the extra time 
 $(\tEc-\tEo)$ unpaired without escaping into lead $3$
 (to close a loop of length $\tEc$).
 The first probability is $\exp[-(2-\rho)\tEo/\tau_{\phi}]$ while the
 second is $\exp[-(\tEc-\tEo)/\tau_{\phi}]$.

 We note that  if we consider a nearly opaque barrier, 
 the relevant time scale involved in the exponent is $\tEc+\tEo \simeq 2 \tEc$.
 Thus by tuning the opacity of the barrier, we can vary the
 exponential contribution to dephasing from $\exp[-\tEc/\tau_\phi]$
 to $\exp[-2\tEc/\tau_\phi]$, but we cannot remove the exponent.
 In particular, we cannot mimic dephasing due to a real environment with
 $\xi \sim L$, Eq.~(\ref{eq:gwl-env}), since it has only a power-law dephasing.

 There is to date no theory for conductance
 fluctuations at finite Ehrenfest time in the presence of tunnel
 barriers, and constructing such a theory would require a formidable 
 theoretical endeavor. Numerically, Ref.~[\onlinecite{Two04.2}] 
 observed that, in contrast to the dephasing lead model 
 without tunnel-barrier, 
 conductance fluctuations in the presence of a dephasing lead
 with a tunnel barrier are exponentially damped $\propto \exp[-\tE/\tau_\phi].$

 %--------------------------------------------------------------------------------------------%
 \subsection{Multiple dephasing leads} 

 The $n$ probe dephasing model consists of adding $n$ fictitious leads to the 
 cavity (labeled $\{3, \cdots, n+2\}$)
 in addition to lead ${\rm L}$, ${\rm R}$.
 The voltage on each supplementary lead is tuned so that 
 the current it carries is zero. 
 Without loss of generality we defined $V_{\rm R}=0$, then
 we get the set of equations 
 \begin{subequations}
 \begin{eqnarray}
 \label{eq:multi-dephasing-a}
 I_{\rm R} &=& T_{\rm RL}V_{\rm L} + {\bf T}_{\rm R}^T {\bf V}  
 \\ 
 \label{eq:multi-dephasing-b}
 0= {\bf I} &=& - {\bf T}_{\rm sub}{\bf V}   + {\bf T}_{\rm L}V_{\rm L}
 \end{eqnarray}
 \end{subequations}
 where the superscript-$T$ indicates the transpose.
 The column-vectors ${\bf I}$, ${\bf V}$ have an $i$th element given by 
 the current or voltage (respectively) for the dephasing lead $i \in \{3,n+2\}$.
 The column-vectors ${\bf T}_{\rm L}$ and ${\bf T}_{\rm R}$ 
 have an $i$th element given by $T_{Li}$ and $T_{Ri}$, respectively. 
 Finally ${\bf T}_{\rm sub}$ has matrix elements given by
 \begin{eqnarray}
 [{\bf T}_{\rm sub}]_{ij} &=& N_i \de_{ij} - T_{ij} \\ 
 &=&\left[ {\textstyle \sum_{k\neq j}}T_{kj}\right] \de_{ij} 
 - T_{ij}(1-\de_{ij}) \, ,
 \end{eqnarray}
 where again $i,j \in \{3,n+2\}$. 
 Substituting ${\bf V}$ from Eq.~(\ref{eq:multi-dephasing-b}) into
 Eq.~(\ref{eq:multi-dephasing-a}) and using $I_{\rm R}=gV_{\rm L}$
 gives us the conductance from 
 ${\rm L}$ to ${\rm R}$ as
  \begin{equation}
 g = T_{\rm LR} + {\bf T}_{\rm L}^T 
 {\bf T}_{\rm sub}^{-1} {\bf T} _{\rm R} \, .
 \end{equation}
 Thus finding $g$ requires the inversion of ${\bf T}_{\rm sub}$.
 This is cumbersome, so instead we present a simple argument to extract 
 only the information we are interested in -- the nature of the
 exponential in the dephasing.

 We argue that 
 whatever the formula
 for conductance for $n$ dephasing leads is, we can expand it in powers of
 $N$ and collect the ${\cal O}[N^0]$-terms to get a formula 
 for weak-localization of the form
 \begin{eqnarray}
 g^{wl} = \delta T_{\rm LR} + \sum_{j=3}^{n+2} A_j  \delta T_{{\rm L}j} 
 + B_j  \delta T_{j{\rm R}} + \sum_{i,j=3}^{n+2} C_{ij} \delta T_{ij}.
 \end{eqnarray}
 Here, the sum is over all dephasing leads.
 To get the prefactors $A_j,B_j, C_{ij}$ we would have solved the full problem
 by inverting ${\bf T}_{\rm sub}$, however we can already see that,
 to leading order,
 they are combinations of Drude conductances and thus
 independent of the Ehrenfest time.  
 In contrast all the weak-localization contributions contain an exponential
 of the form [$\tau_{\rm D_1}$ and $\tau_{\rm D_2}$ are defined
 below Eq.~(\ref{gijwl})]
 \begin{eqnarray}
 \exp [-\tEo/\tau_{\rm D_2} +(\tEc-\tEo)/\tau_{\rm D_1}] \,.
 \end{eqnarray}
 Thus defining $\tau_\phi^{-1}$ as the rate of escaping into any of the
 dephasing leads, so $\tau_\phi^{-1} = (\tau_0 L)^{-1} \sum_{j=3}^n \rho_j W_j$,
 we see that 
 $g^{\rm wl}$ decays with an exponential 
 \begin{eqnarray}
 g^{\rm wl} \propto \exp [-(1-\tilde\rho)\tEo/\tau_{\phi} -  \tEc/\tau_{\phi}],
 \end{eqnarray}
 where we define $\tilde \rho$ such that 
 $\tilde{\rho} \tau_{\phi}^{-1} = (\tau_0L)^{-1}\sum_j \rho_j^2 W_j$.
 We have just shown that multiple dephasing leads cause
 an exponential suppression of the weak-localization which is qualitatively
 similar to that caused by a single dephasing lead. 
 The exponent is proportional to the Ehrenfest time,
 and contains no independent parameter analogous to $\xi$.

 %~~~~~~~~~~~~~~~~~~~~~~~~~~~~NUMERICAL SIMULATIONS~~~~~~~~~~~~~~~~~~~~~~~%
 \section{NUMERICAL SIMULATIONS}\label{NUM-SEC}

 \subsection{Open kicked rotators}

 Because of the slow, logarithmic increase of $\tE$ with 
 the size $M$ of the Hilbert space, the ergodic semiclassical regime 
 $\tau_{\rm E} \gtrsim \tau_{\rm
 D}$, $\lambda \tau_{\rm D} \gg 1$ is unattainable by standard
 numerical methods. We therefore follow Refs.~[\onlinecite{Andreev,Two03,JacqSukho04,Two04,Two04.2,Bro06-quasi-wl}] and
 consider the open kicked rotator model.

 \subsubsection{Dephasing lead model}

 The system is described by the time-dependent Hamiltonian 
 \begin{eqnarray}\label{krot}
 H = \frac{(p-p_0)^2}{2} + K \cos(x-x_0) \sum_n \delta(t-n \tau_0).
 \end{eqnarray}
 The kicking strength $K$ drives the dynamics from
 integrable ($K=0$) to fully chaotic ($K\agt 7$). The Lyapunov exponent
 in the classical version of the kicked rotator is given by
 $\lambda \tau_0 \approx\ln (K/2)$. 
 In most quantum simulations,
 however, one observes an effective Lyapunov 
 exponent $\lambda_{\rm eff}$ instead that is systematically 
 smaller than $\lambda$ by as much as 30\%~\cite{Sil03}.
 Two parameters, $p_0$ and $x_0$, are 
 introduced to break two parities and drive the crossover
 from the $\beta=1$ to the $\beta=2$ universality class \cite{Meh91},
 corresponding to breaking the time reversal symmetry. 

 We quantize the Hamiltonian of Eq.~(\ref{krot}) on the torus $x,p \in[0,2 \pi]$, by discretizing the 
 momentum coordinates as $p_l=2 \pi l/M$, $l=1,\ldots M$. 
 A quantum representation of the Hamiltonian of Eq.~(\ref{krot}) is
 provided by the unitary $M \times M$ 
 Floquet operator $U$, 
 which gives the time evolution for one iteration of the map. 
 For our specific choice of the kicked rotator,
 the Floquet operator has matrix elements
 \begin{eqnarray}\label{kickedU}
 U_{l,l'} &=& M^{-1/2} e^{-(\pi i/2 M) [(l-l_0)^{2}+(l'-l_0)^2]}
 \\
 & \times & \sum_m e^{2 \pi i m(l-l')/M}   
 e^{-(iMK/2\pi) \cos[2\pi (m-m_0)/M]}, \nonumber
 \end{eqnarray}
 with $l_0=p_0 M/2 \pi$ and $m_0=x_0 M/2 \pi$. The Hilbert space
 size $M$ is given by the ratio $L/\lambda_{\rm F}$ of the linear system size to the Fermi wavelength.

 To investigate transport, we open the system by
 defining $n+2 \ge 2$ contacts to 
 leads via absorbing phase-space strips
 $[l_i-N_i/2,l_i+N_i/2-1]$, $i=1,2 \ldots n+2$. With 
 $N_{\rm T} = \sum_i N_i$, we construct a $N_{\rm T} \times N_{\rm T}$ scattering matrix 
 from the Floquet operator $U$ as \cite{Fyo00}
 \begin{equation}\label{smatrix}
 S(\varepsilon) = \sqrt{{\mathcal I} -{\mathcal P} {\mathcal P}^\dagger } - {\mathcal P} \Big[e^{-i \varepsilon} \, {\mathcal I} - U \sqrt{{\mathcal I}-{\mathcal P}^\dagger {\mathcal P}} \Big]^{-1} U {\mathcal P}^\dagger.
 \end{equation}
 The $N_{\rm T} \times M$ projection matrix ${\mathcal P}$, which 
 describes the coupling to the leads, has matrix elements
 \begin{eqnarray}\label{lead}
  {\mathcal P}_{l,m}=\left\{\begin{array}{ll}
 \gamma_l \, \delta_{lm} & \mbox{if $l \in \bigcup_i  \; \{l^{(i)} \}$},\\
 0& \mbox{otherwise,}
 \end{array}\right.
 \end{eqnarray}
 where $|\gamma_l|^2 = \rho_l \in [0,1]$ gives the transparency of the contact to the external
 channel $l$, and $\bigcup_i  \; \{l^{i} \}$ denotes the ensemble of cavity modes coupled
 to the external leads. Below we focus on perfectly transparent contacts, 
 $\rho_l = \gamma_l = 1$, $\forall l$. 

 The conductance (\ref{3leadmodelDr}) in the dephasing lead model is obtained 
 from Eq.~(\ref{smatrix}) with 
 $n+2 = 3$ leads. 
 The transport leads carry $N = N_{\rm L} = N_{\rm R}$ channels, which defines
 the dwell time through the system as $\tD/\tau_0 = M/2 N$. The dephasing time 
 $\tau_\phi/\tau_0 = M/N_3$ is
 defined by the number $N_3$ of channels carried by the third, dephasing lead,
 and the Ehrenfest time is given by
 \begin{equation}\label{taueorder1}
 \tEc =\lambda^{-1}\left[\ln{M}+{\cal{O}}(1)\right].
 \end{equation}

 \subsubsection{The open kicked rotator coupled to an environment}

 We extend the kicked rotator model to account for the coupling to external degrees of freedom.
 The exponential increase of memory size with number of particles forces us to focus on 
 an environment modeled by a single chaotic particle. Therefore, we follow Ref.~[\onlinecite{Pet06}] and
 consider two coupled kicked rotators (with $i = {\rm sys},  \, {\rm env}$)
 \begin{eqnarray}\label{krotcoupled}
 && {\cal H} = H_{\rm sys} + H_{\rm env} + {\cal U}, \nonumber \\
 && H_{i}   =  \frac{(p_i-p_0)^2}{ 2} + K_{i} \cos(x_{i}-x_0) 
  \sum_n \delta(t-n\tau_{0}),\nonumber\\
 &&{\cal U}  = \varepsilon \; \sin (x_{\rm sys}-x_{\rm env}-0.33 )  \sum_n \delta(t-n\tau_{0}).\,
 \end{eqnarray}
 In this model, the interaction potential ${\cal U}$ acts at the same time as the kicks, which
 facilitates the construction of the $S$-matrix. For this particular choice
 of interaction, the correlation length $\xi=L$ so that one expects a
 universal behavior of dephasing,
 as in Eq.~(\ref{eq:dephasing-wl-without-Ehrenfest}).
 The quantum representation of the coupled Hamiltonian is a
 unitary $(M_{\rm sys } \,M_{\rm env })  \times (M_{\rm sys } \,M_{\rm env }) $ Floquet operator.
 We open the system (and not the environment)
 to two external leads by means of extended projectors 
 ${\mathbb P}_{\rm tot}  = P^{\rm (L)}\otimes   
 {\mathbb I}_{\rm env}+P^{\rm (R)}\otimes  {\mathbb I}_{\rm env}$. A 
 straightforward generalization of Eq.~(\ref{smatrix}) 
 defines a $(N_{\rm T} M_{\rm env}) \times
 (N_{\rm T} M_{\rm env}) $ extended scattering matrix, from which we evaluate the
 conductance via Eq.~(\ref{conductance}).
 We focus on the symmetric situation where the two leads carry the same number $N$ of channels and average our data over a set of 
 pure but random initial environment density matrices 
 $\eta_{\rm env}({\bf q},{\bf q}';t=0)$. We estimate the dephasing time
 from the $\varepsilon$-induced broadening of two-particle levels in the
 corresponding closed two-particle kicked rotator, $\tau_\phi^{-1} = 0.43 
 (\varepsilon/\hbar_{\rm eff})^2$, with $\hbar_{\rm eff} = 2 \pi/M$,
 the effective Planck constant~\cite{Pet06}.

  %%%%%%%%%%%%%%%%%%%%%%%%%%%%%%%%%%%%%%%%% %%%%%%%
 \begin{figure}
 \begin{center}
 \includegraphics[width=8.0cm]{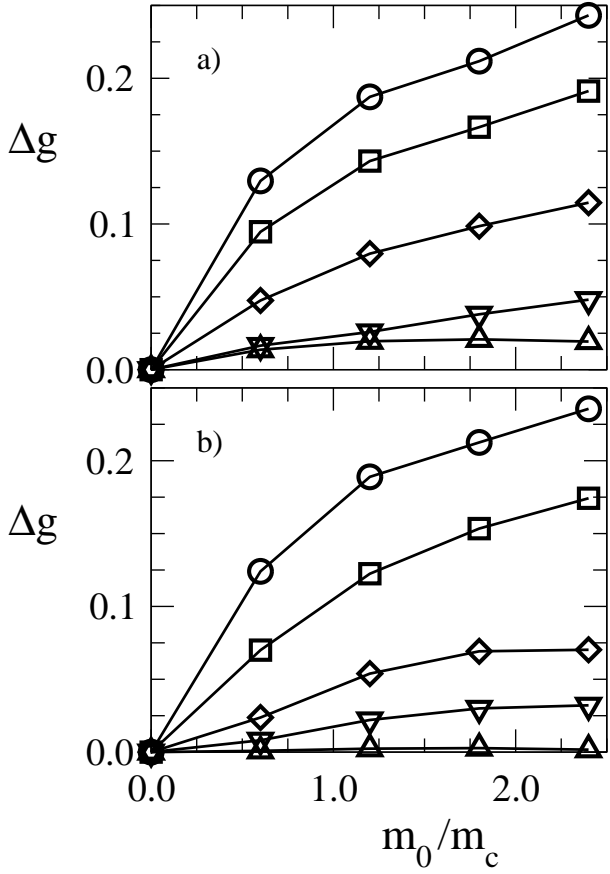}
 \end{center}
 \caption{\label{fig:weakloc_num} a) Magnetoconductance curves 
 $\Delta g(x_0)=g(x_0)-g(0)$ for the double kicked rotator model (see text) with
 $K_{\rm sys} = K_{\rm env}=34.08$ 
 ($\lambda_{\rm eff} \approx 2$), 
 $\tau_{\rm D}/\tau_{0}=8$, $\xi/L = 1 $ and Hilbert space sizes
 $M_{\rm sys}=256$, $M_{\rm env}=16$. Different symbols correspond to different dephasing
 times $\tau_\phi/\tD = \infty$ ($\hbar_{\rm eff}^{-1}\varepsilon  =0 $, circles), $\tau_\phi/\tD = 4.8$ ($\hbar_{\rm eff}^{-1}\varepsilon  \simeq 0.25 $, squares),  $\tau_\phi/\tD =1.2$
 ($\hbar_{\rm eff}^{-1}\varepsilon  \simeq  0.5 $, diamonds),  $\tau_\phi/\tD =0.3$
 ($\hbar_{\rm eff}^{-1}\varepsilon  \simeq 1$, downward triangles) and
  $\tau_\phi/\tD =0.07$ ($\hbar_{\rm eff}^{-1}\varepsilon  \simeq 2$, upward triangles). Data are
 averaged over $25$ different lead positions, each with $25$ different quasi-energies and $10$ different initial environment states. b) Magnetoconductance curves for the open kicked rotator
 with transparent dephasing lead (see text) with $K=34.08$, $\tau_{\rm D}/\tau_{0}=8$ and Hilbert
 space size $M=256$. Different symbols correspond to different dephasing
 times $\tau_\phi/\tD = \infty$ (circles), $\tau_\phi/\tD = 5$ (squares),  $\tau_\phi/\tD =1.25$
 (diamonds),  $\tau_\phi/\tD =0.5$
 (downward triangles) and
  $\tau_\phi/\tD =0.25$ (upward triangles). Data are
 averaged over $225$ different lead positions, each with $50$ different quasi-energies.}
 \end{figure}
 %%%%%%%%%%%%%%%%%%%%%%%%%%%%%%%%%%%%%%%%%%%%%%%%

 %%%%%%%%%%%%%%%%%%%%%%%%%%%%%%%%%%%%%%%%%%%
 \begin{figure}
 \begin{center}
 \psfrag{GG}{$\mbox{ \hspace{-1.cm} \Large$\delta g(2.4)-\delta g(0)$}$}
 \psfrag{tt}{$\mbox{ \hspace{-1.cm} \vspace{0.5cm}\Large$\tD/\tau_\phi$}$}
 \includegraphics[width=8cm]{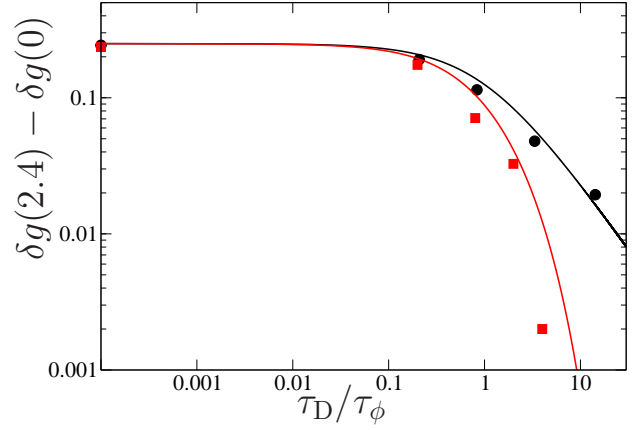}
 \end{center}
 \caption{\label{dg(2.4)} (Color online) Amplitude 
 $\delta g(m_0/m_c=2.4)-\delta g(m_0/m_c=0)$
 of the weak localization correction to the conductance as a function of 
 $\tD/\tau_\phi$ for the double kicked rotator model (circles) 
 and the open kicked rotator with transparent dephasing lead 
 (squares).
 The black line gives the universal algebraic behavior
 $(1+\tD/\tau_\phi)^{-1}/4$, 
 and the red line 
 is a guide to the eye, including an exponential decay with $\tEc = 2.78$,
 on top of the universal decay.
 }
 \end{figure}

 \subsection{Weak-localization with dephasing}

 To investigate weak-localization, 
 we follow the same procedure as in Ref.~[\onlinecite{Bar05}] of taking
 a constant nonzero $p_0$, while varying $x_0$ (which hence plays the role of
 a magnetic field). The obtained magnetoconductance in the absence of
 dephasing is Lorentzian~\cite{Jac06,Bar05},
 \begin{eqnarray}\label{eq:gwl-lorentz}
 g^{\rm wl}(\Phi)   &=& {g^{\rm wl}(0) \over 1 + (m_0/m_c)^2 } \;,
 \end{eqnarray}
 with $m_c=4 \pi/K \; \sqrt{M \tD}$.
 In Fig.~\ref{fig:weakloc_num} we compare the suppression of weak-localization
 for the system-environment kicked rotator [panel a), top] and
 the dephasing lead kicked rotator [panel b), bottom]. 
 For both models, we show five magnetoconductance curves,
 corresponding to five different ratios $\tau_\phi/\tD$. All curves exhibit
 the expected Lorentzian behavior vs. $m_0/m_c$, however, 
 the amplitude
 of the magnetoconductance is reduced as $\tau_\phi/\tD$ is reduced. This
 allows one to extract the $\tau_\phi$-dependence of $g^{\rm wl}$. For the
 system-environment model, we found no significant departure from our
 analytical prediction, Eq.~(\ref{eq:gwl-env}) with $\tau_\xi=0$ (since the
 interaction in Eq.~(\ref{krotcoupled}) has $\xi=L$). The same behavior
 is observed for the dephasing lead model, as long as $\tau_\phi/\tD$
 is large, however, compared to the system-environment kicked rotator,
 the damping of magnetoconductance accelerates
 as $\tau_\phi/\tE$ becomes comparable to or smaller than one. When
 this regime is reached, magnetoconductance curves for the dephasing lead
 model lies significantly below those of the system-environment
 model, even when the latter corresponds to shorter dephasing times
 (compare
 in particular the upward triangles in both panels of 
 Fig.~\ref{fig:weakloc_num}). This behavior is further illustrated in
 Fig.~\ref{dg(2.4)}, where we plot the amplitude
 $\delta g(m_0/m_c=2.4)-\delta g(m_0/m_c)$ of the weak localization
 corrections to the conductance as a function of $\tD/\tau_\phi$.
 The data for the double kicked rotator nicely line up on the universal
 algebraic behavior $(1+\tD/\tau_\phi)^{-1}/4$ without any fitting parameter.
 This is clearly not the case for the dephasing lead model, where an additional,
 exponential dependence on $\tau_\phi$ emerges. We attribute this to the
 exponential damping factor $\propto \exp[-\tE/\tau_\phi]$ of 
 Eq.~(\ref{eq:weakloc_3lead}). 
 The data presented in Figs.~\ref{fig:weakloc_num} and \ref{dg(2.4)} 
 for $\tau_\phi^{-1}=0$ exhibit a weak dependence on $\exp[-\tEc/\tD]$ only, 
 with $\tEc \lesssim 0.8$, which we attribute to terms of order one in
 Eq.~(\ref{taueorder1}).

 All collected data (including some that we do not present here) thus
 confirm qualitatively -- if not quantitatively -- 
 the validity of Eq.~(\ref{eq:weakloc_3lead}) for the dephasing lead model.

 \subsection{Conductance fluctuations in the dephasing lead model}

\begin{figure}
 \begin{center}
 \includegraphics[width=8cm]{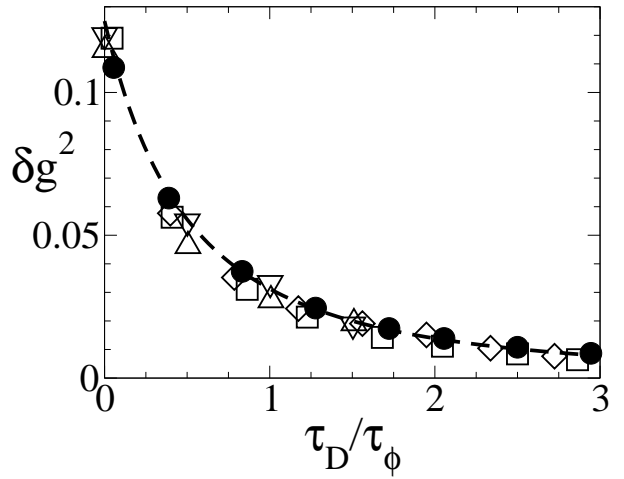}
 \end{center}
 \caption{\label{fig:ucf}  Variance of the conductance vs. $\tD/\tau_\phi$ for the open kicked rotator
 with $K=14.$ and $\tD/\tau_0=5$ (empty symbols), transparently coupled to a dephasing lead.   
 Different symbols correspond to different Hilbert space sizes (and hence different
 $\tEc$) $M=128$ (squares, $\tEc/\tD=0.6$),
 $M=512$ (diamonds, $\tEc/\tD=0.75$), $M=2048$ (upward triangles, $\tEc/\tD=0.9$) and 
 $M=8192$ (downward triangles, $\tEc/\tD=1.1$). Additional data for $K=144.$, $\tD=25$ and
 $M=2048$ are also shown (full circles, $\tEc/\tD=0.08$).
 The dashed line shows the universal behavior
 of Eq.~(\ref{eq:dephasing-ucf-without-Ehrenfest}). Unlike for weak-localization (see Fig.~\ref{fig:weakloc_num})
 and for the dephasing lead model with partial transparency~\cite{Two03}, the 
 behavior of $\delta g^2$
 remains universal and shows no noticeable dependence on $\tEc/\tD$. Data are averaged over
 50 different quasi-energies and from 50 (for $N=8192$) to 500 (for $N=128$ and 512)
 different lead positions.}
 \end{figure}

 Conductance fluctuations were 
 studied numerically in Ref.~[\onlinecite{Two04.2}] for the 
 dephasing lead model with a tunnel barrier of low transparency, and an
 exponential damping ${\rm var} (g) \propto \exp[-2\tE/\tau_\phi]$ was reported.
 Instead, here we consider a model in which the dephasing lead 
 is transparently coupled 
 to the system. In Fig.~\ref{fig:ucf},
 we show data for ${\rm var} (g)$ against $\tau_\phi/\tD$, which 
 is varied only by varying the width of the third, dephasing lead.
 There are four data sets (empty symbols) corresponding to a
 fixed classical configuration at 
 different stages in the quantum-classical crossover, i.e. 
 with increasing ratio $M=L/\lambda_{\rm F} \in [128,8192]$.
 As $M$ increases, so does $\tE/\tau_\phi$, however no
 change of behavior of ${\rm var} (g)$ is observed. These data are
 compared to a fifth set obtained in the universal regime,
 $\tE/\tD \ll 1$, and the universal prediction of 
 Eq.~(\ref{varg_double_lorentzian})
 (dashed line). These data confirm our analytical result, 
 Eq.~(\ref{varg_double_lorentzian}),
 that ${\rm var} (g)$ exhibits no Ehrenfest time dependence for  
 the dephasing lead model with perfectly transparent contacts.

 %%%%%%%%%%%%%%%%%%%%%%%%%%%%%%%%%%%%%%%%%%%

 %~~~~~~~~~~~~~~~~~~~~~~~~~~CONCLUSIONS ~~~~~~~~~~~~~~~~~~~~~~~~~~~~~%
 \section{CONCLUSIONS}\label{CONCL-SEC}

 We have investigated the dephasing properties of  open quantum chaotic 
 system, focusing on the deep semiclassical limit where the Ehrenfest
 time is comparable to or larger than the dwell time through the system.
 We treated three models of dephasing. In the first one, the transport
 system is capacitively coupled to an external quantum chaotic system.
 For that model, we developed a new scattering formalism, based
 on an extended scattering matrix ${\mathbb S}$, 
 including the degrees of freedom of
 the environment. Transport properties are extracted from 
 ${\mathbb S}$, once the environment has been traced out properly. 
 In that model, we find that, in addition to the universal algebraic
 suppression $g^{\rm wl} \propto (1+\tau_{\rm D}/\tau_\phi)^{-1}$
 with the dwell time $\tau_{\rm D}$ through the cavity 
 and the dephasing rate $\tau_\phi^{-1}$, weak-localization
 is exponentially suppressed by a factor 
 $\propto \exp[-\tau_\xi/\tau_\phi]$, with a new time scale
 $\tau_\xi$ depending on the correlation length of the coupling potential
 between the system and the environment.

 The second model we treated is that of dephasing due to a classical noise
 field.  We show that the new time scale $\tau_\xi$ 
 plays the same role here as in the system-environment model.
 We then consider a classical Johnson-Nyquist noise model of electron-electron
 interactions. We show that 
 $\xi \sim \lambda_{\rm F}$ (and so
 $\tau_\xi$ equals the Ehrenfest time) 
 when dephasing is dominated by electron-electron interactions 
 within the system,
 but that $\xi \sim D$ when dephasing is dominated by
 interactions between electrons in the system and those in a gate, 
 a distance $D$ away.

 The third model we treated is the dephasing lead model. We found a similar
 exponential suppression of weak-localization. To our surprise, however,
 it is the Fermi wavelength, not the dephasing-lead's width,  
 which plays a role similar to $\xi$ in that model.
 This inequivalence between the dephasing lead model and 
 dephasing due to a real environment or classical noise
 can be most clearly seen in a situation where the interaction with a 
 real environment has $\xi \simeq L$.
 Then the environment induces only power-law dephasing,
 which is impossible to mimic with a dephasing lead. For the
 dephasing lead model, we also showed analytically and numerically
 that conductance fluctuations exhibit only power-law dephasing
 if the connection between the cavity and the dephasing lead
 is perfectly transparent. This is to be contrasted with the 
 exponential dephasing observed for the dephasing lead model with
 tunnel barrier, and reflects the fact that the presence of
 tunnel barriers violates a sum rule otherwise 
 preserving the universality of conductance
 fluctuations vs. $\tE/\tD$~\cite{Bro06-ucf}. 

 Related results have been obtained for conductance fluctuations
 in Ref.~[\onlinecite{Bro06-quasi-ucf}], where different behaviors have
 been predicted for external sources -- which give 
 similar dephasing as our environment model -- 
 and internal source of dephasing -- which qualitatively
 reproduce the prediction of Ref.~[\onlinecite{Ale96}].

 %~~~~~~~~~~~~~~~~~~~~~~~~~~ACKNOWLEDGMENTS ~~~~~~~~~~~~~~~~~~~~~~~~~~~~~%
 \section*{ACKNOWLEDGMENTS}

 We thank I. Aleiner, P. Brouwer, M. B\"uttiker and M. Polianski
 for useful and stimulating discussions. While working on this project,
 C. Petitjean was supported by the Swiss National Science Foundation,
 which also funded
 the visits to Geneva of P.Jacquod and R.Whitney. 
 P. Jacquod expresses his gratitude to M. B\"uttiker and the
 Department of Theoretical Physics at the University of Geneva for their
 hospitality. This work was initiated in the summer of 2006
 at the Aspen Center for Physics.

 \appendix

 \begin{widetext}

 \section{Scattering approach to transport in the presence of an environment}
 \label{extendS}

 Here we extend the scattering approach to transport to account
 for those environmental degrees of freedom which couple to the system
 being studied. 
 We follow the lines of 
 the derivation of the expression for noise presented in 
 Ref.~[\onlinecite{Blanter-review}], focusing on the two-terminal configuration.
 The following derivation is valid in the limit of pure dephasing, when there is
 no energy/momentum exchange between system and environment.

 \subsection{Current operator and conductance}
 \label{sect:current-operator}

 In the presence of an environment, 
 the current operator at time $t$ on a cross-section deep inside of 
 lead $\alpha=L,R$ (where there is no system-environment interaction) 
 reads
 \begin{eqnarray}
 \hat{I}_\alpha (t) &=& \frac{e}{h} \int {\rm d}E \;
 {\rm d}E' e^{i (E-E') t/\hbar}
 \sum_{n \in \alpha} \Big[\hat{a}_{\alpha n}^\dagger(E') \hat{a}_{\alpha n}(E)
 - \hat{b}_{\alpha n}^\dagger(E') \hat{b}_{\alpha n}(E) \Big] \times {\mathbb I}_{\rm env}.
 \end{eqnarray}
 The second quantized
 operators $\hat{a}^{(\dagger)}$ and $\hat{b}^{(\dagger)}$ create and 
 destroy incoming and outgoing system particles
 respectively.  Since the environment particles carry no current, the 
 current operator acts as
 the identity operator ${\mathbb I}_{\rm env}$ in the environment sub-space.
 We could write ${\mathbb I}_{\rm env}$ in terms of second quantized
 operators, however for our present purpose, 
 it is more convenient to write it as 
 \begin{eqnarray}
 {\mathbb I}_{\rm env} &=& \int \rmd {\bf q} |{\bf q}\rangle \langle {\bf q}| .
 \end{eqnarray}

 As in Ref.~[\onlinecite{Blanter-review}] we now back evolve the outgoing-states into incoming-states, this time with an $S$-matrix 
 which also depends on the coordinates of the environment,
 \begin{equation}\label{extended-S}
 \hat{b}_{\alpha n}(E)  \langle {\bf q}|  = 
 \sum_{\beta; j} \int {\rm d} {\bf q}_0 \;
 {\mathbb S}_{\alpha \beta; n j}({\bf q},{\bf q}_0;E,\tilde E) \,
 \hat{a}_{\beta j}(\tilde E) \, \langle {\bf q}_0|.
 \end{equation}
 Here, ${\mathbb S}_{\alpha \beta; n j}({\bf q},{\bf q}_0;E, \tilde E)$ gives the
 transmission amplitude from channel $j$ in lead $\beta$ with energy
 $\tilde E$ to channel $n$
 in lead $\alpha$ with energy $E$, 
 while simultaneously, the environment evolves from ${\bf q}_0$ to
 ${\bf q}$.
 We can set $\tilde E =  E$, because throughout this article
 we only consider the regime of pure dephasing. 
 Using (\ref{extended-S}) we rewrite
 the current operator as
 \begin{eqnarray}\label{eq:current-op}
 \hat{I}_\alpha (t) &=& \frac{e}{h}\int \rmd {\bf q} \int {\rm d}E \;
 {\rm d}E' e^{i (E-E') t/\hbar} \; \sum_{n} 
 \Bigg[ 
 \hat{a}_{\alpha n}^\dagger(E') \; \hat{a}_{\alpha n}(E) \;
 \langle \Psi_{\rm env}|{\bf q}\rangle \langle {\bf q} |\Psi_{\rm env}\rangle 
 \nonumber \\
 && \qquad 
 -\int {\rm d} {\bf q}'_0 \; {\rm d}{\bf q}_0
 \sum_{\gamma,\beta} \sum_{jk} 
 \Big({\mathbb S}_{\alpha \gamma; nj}({\bf q},{\bf q}'_0)\Big)^\dagger \;
 {\mathbb S}_{\alpha \beta; n k}({\bf q},{\bf q}_0) \;
 \hat{a}_{\gamma j}^\dagger(E') \; \hat{a}_{\beta k}(E) \;
 \langle \Psi_{\rm env}|{\bf q}'_0 \rangle 
 \langle {\bf q}_0 |\Psi_{\rm env}\rangle 
 \Bigg].
 \end{eqnarray}
 We next rewrite the first line of Eq.~(\ref{eq:current-op}) as
 \begin{eqnarray}
 & & \hskip -6mm
 \int \rmd {\bf q} \sum_n 
 \hat{a}_{\alpha n}^\dagger(E') \hat{a}_{\alpha n}(E) 
 \langle \Psi_{\rm env}|{\bf q}\rangle 
 \langle {\bf q} |\Psi_{\rm env}\rangle
 \nonumber \\
 &=& \int \rmd {\bf q}_0 \rmd {\bf q}'_0 \sum_{\gamma,\beta} \sum_{jk} 
 \delta({\bf q}'_0-{\bf q}_0) 
 \delta_{\gamma\alpha} \delta_{\beta\alpha} \delta_{jk} 
 \hat{a}_{\gamma j}^\dagger(E') \hat{a}_{\beta k}(E) 
 \langle \Psi_{\rm env}|{\bf q}'_0\rangle 
 \langle {\bf q}_0 |\Psi_{\rm env}\rangle.
 \label{eq:delta-functions}
 \end{eqnarray}
 Finally, we write the current operator 
 in terms of the initial environment
 density-matrix
 $\eta_{\rm env}({\bf q}_0,{\bf q}'_0)=
 \langle {\bf q}_0 |\Psi_{\rm env}\rangle
 \langle \Psi_{\rm env}|{\bf q}'_0\rangle$,
 \begin{eqnarray}\label{current-operator}
 \hat{I}_\alpha (t) &=& \int \rmd {\bf q}'_0\rmd {\bf q}_0 
 \hat{I}_\alpha^{\rm (red)}({\bf q}'_0,{\bf q}_0;t)
 \eta_{\rm env}({\bf q}_0,{\bf q}'_0),
 \end{eqnarray}
 where 
 we defined the reduced current operator as
 \begin{subequations}
 \begin{eqnarray}\label{reduced-current-operator}
 \hat{I}_\alpha^{\rm (red)} ({\bf q}'_0,{\bf q}_0 ;t) 
 &=& \frac{e}{h}\int \rmd {\bf q} 
 \int {\rm d}E \; {\rm d}E' e^{i (E-E')t/\hbar} \; 
 \sum_{\gamma,\beta} \sum_{j,k} 
 B_{\gamma\beta}^{jk}(\al,E,E'; {\bf q}'_0,{\bf q}_0)
 \hat{a}_{\gamma j}^\dagger(E') \; \hat{a}_{\beta k}(E) ,
 \\
 \label{Btensor}
 B_{\gamma\beta}^{jk}(\al,E,E'; {\bf q}'_0,{\bf q}_0) 
 &=&  
 \delta({\bf q}'_0-{\bf q}_0) 
 \delta_{\gamma\alpha} \delta_{\beta\alpha} \delta_{jk} 
 - \int \rmd {\bf q} \sum_n
 \Big({\mathbb S}_{\alpha \gamma; nj}({\bf q},{\bf q}'_0)\Big)^\dagger \;
 {\mathbb S}_{\alpha \beta; n k}({\bf q},{\bf q}_0). 
 \end{eqnarray}
 \end{subequations}
 The current is obtained by taking the expectation value of the current operator
 over the system, using
 \begin{eqnarray}
 \doublelangle 
 \hat{a}_{\gamma j}^\dagger(E') \; \hat{a}_{\beta k}(E) \doublerangle &=&
 \; \delta_{\gamma \beta} \; \delta_{jk} \; \delta(E-E') \; f_\beta (E)
 \label{operator-contractions}
 \end{eqnarray}
 where $f_\beta$ is the Fermi function in lead $\beta$.
 The unitarity of ${\mathbb S}$ implies,
 \begin{equation}
 \int {\rm d}{\bf q} \sum_{\delta, n} 
 \Big({\mathbb S}_{\delta \gamma; nj}({\bf q},{\bf q}'_0)\Big)^\dagger \;
 {\mathbb S}_{\delta \beta; n k}({\bf q},{\bf q}_0) \;
 = 
 \delta({\bf q}'_0 - {\bf q}_0) \delta_{\beta\gamma} \delta_{jk} .
 \label{eq:unitarity}
 \end{equation}
 We use this latter equality 
 to rewrite the Kronecker $\delta$'s in Eq.~(\ref{operator-contractions}).
 Finally the current in the left lead is
 \begin{equation}\label{current}
 \doublelangle I_{L} \doublerangle = \frac{e}{h} \sum_{n,k} \int 
 {\rm d}{\bf q}'_0{\rm d}{\bf q}_0 {\rm d}{\bf q} \int {\rm d}E 
 \Big({\mathbb S}_{LR; n k}({\bf q},{\bf q}'_0)\Big)^\dagger \;
 {\mathbb S}_{LR; n k}({\bf q}, {\bf q}_0)
 \Big[f_{L}(E)-f_{R}(E) \Big] \eta_{\rm env}({\bf q}_0, {\bf q}'_0).
 \end{equation}
 In the limit of zero temperature in the leads, and assuming that the scattering
 matrix is not too strongly energy dependent, Eq.~(\ref{current})
 leads to the linear conductance
 \begin{equation}\label{conductance}
 G = \frac{e^2}{h} \sum_{n,k} \int 
 {\rm d}{\bf q}'_0 {\rm d}{\bf q}_0 {\rm d}{\bf q} 
 \Big({\mathbb S}_{LR; n k}({\bf q}, {\bf q}'_0)\Big)^\dagger \;
 {\mathbb S}_{LR; k n}({\bf q}, {\bf q}_0) 
 \eta_{\rm env}({\bf q}_0, {\bf q}'_0),
 \end{equation}
 with scattering matrices to be evaluated at the Fermi energy.
 From Eqs~(\ref{current}) and (\ref{conductance}), 
 we see that both current and conductance are
 obtained by tracing over the environmental degrees of freedom of the 
 square of the extended scattering matrix. Besides this prescription, 
 these two equations are 
 extremely similar to their counterpart in the standard scattering approach to
 transport. We also note that conductance fluctuations can be obtained by
 squaring Eqs.~(\ref{current}) and (\ref{conductance}).

 It is legitimate to expect that a complex environment -- such as the chaotic 
 system considered in this article --
 has a complicated initial wavefunction, which, under ensemble averaging,
 is uncorrelated with itself on all scales greater than the environment 
 wavelength. This justifies us treating the initial environment state as
 $\langle \eta_{\rm env}({\bf q}_0,{\bf q}'_0)\rangle 
 \sim (2\pi\hbar)^{Nd} \, \delta({\bf q}_0-{\bf q}'_0)/\Xi_{\rm env} $.

 \subsection{Current noise}
 \label{appendix:current-noise}

 We follow similar steps as in the previous Section to
 calculate the zero-frequency current noise. 
 However now we have two current operators, 
 and hence two creation and two annihilation operators for the system
 \cite{Blanter-review}.
 The environment has two ${\mathbb I}_{\rm env}$ operators, each of which 
 we write
 in the form $\int \rmd {\bf q} |{\bf q}\rangle \langle{\bf q}|$
 to get the current time-correlator as
 \begin{eqnarray}
 \label{correl}
 \bdoublelangle \hat I_\beta(t) \hat I_\alpha(0) \bdoublerangle= 
 \Bggdoublelangle 
 \int \rmd {\bf q}_{03} \rmd {\bf q}_{01} \rmd {\bf q}'_{01} \,
 \hat{I}_\beta^{(\rm red)} ({\bf q}'_{01},{\bf q}_{03};t) \,
 \hat{I}_\alpha^{(\rm red)} ({\bf q}_{03},{\bf q}_{01};0) \,
 \eta_{\rm env}({\bf q}_{01},{\bf q}'_{01}) \,
 \Bggdoublerangle.
 \end{eqnarray}
 The reduced current operator 
 $\hat{I}_\beta^{(\rm red)} ({\bf q}',{\bf q};t)$ is given in
 Eq.~(\ref{reduced-current-operator}).
 The zero-frequency noise power is obtained from the
 product of the deviations from the average current 
 at times $0$ and $t$.
 Consequently it is proportional to 
 \begin{eqnarray}
 \int \rmd t \int \rmd {\bf q}_{03} \rmd {\bf q}_{01} \rmd {\bf q}'_{01} 
 \Bdoublelangle
 \big( \, \hat{I}^{\rm (red)}({\bf q}'_{01},{\bf q}_{03};t)
 -\doublelangle 
 \hat I^{\rm (red)}({\bf q}'_{01},{\bf q}_{03};t) \doublerangle \,\big)
 \big( \, \hat{I}^{\rm (red)}({\bf q}_{03},{\bf q}_{01};0)
 -\doublelangle 
 \hat I^{\rm (red)}({\bf q}_{03},{\bf q}_{01};0) \doublerangle \,\big)
 \Bdoublerangle
 \nonumber \\
 \times \eta_{\rm env}({\bf q}_{01},{\bf q}'_{01}) \qquad 
 \end{eqnarray}
 There is only one trace over the environment here
 because we assume we measure the current as a function of time
 in a given experiment (with a given initial $\eta_{\rm env}$),
 and then extract the average current (and the deviations from it)
 from that data set.

 We need to take the following expectation value of products of
 creation/annihilation operators over the system\cite{Blanter-review}
 \begin{eqnarray}
 \bdoublelangle \hat{a}_{\beta m}^\dagger(E_2) \; \hat{a}_{\gamma n}(E_1) 
 \hat{a}_{\beta' m'}^\dagger(E_4) \; \hat{a}_{\gamma' n'}(E_3) \bdoublerangle 
 -
 \bdoublelangle 
 \hat{a}_{\beta m}^\dagger(E_2) \; \hat{a}_{\gamma n}(E_1) \bdoublerangle
 \bdoublelangle 
 \hat{a}_{\beta' m'}^\dagger(E_4) \; \hat{a}_{\gamma' n'}(E_3) \bdoublerangle =
 \nonumber \\
 \delta_{\beta \gamma'} \, \delta_{\gamma \beta'} \,\delta_{m n'} \,
 \delta_{n m'} \, \delta(E_2-E_3) \, \delta(E_1 - E_4)
 f_\alpha (E_2) [1 \mp f_\beta (E_1)],
 \end{eqnarray}
 where the minus (plus) sign stands for fermions (bosons). From this
 we finally get
 the zero-frequency noise power
 \begin{eqnarray}\label{noise}
 S_{\alpha \alpha'}(0) &=& \frac{e^2}{h} \int {\rm d}E \; 
 \int \rmd {\bf q}_{03} \rmd {\bf q}_{01} \rmd {\bf q}'_{01} 
 \sum_{\beta,\gamma}
 \sum_{m,n} B_{\beta\gamma}^{nm}(\alpha,E,E;{\bf q}'_{01},{\bf q}_{03}) 
 B_{\gamma\beta}^{mn}(\alpha',E,E;{\bf q}_{03},{\bf q}_{01}) \,
 \nonumber \\
 & & \times
 \Big( f_\beta(E) [1\mp f_\gamma(E)] + [1\mp f_\alpha(E)] f_\beta(E) \Big)
 \ \eta_{\rm env}({\bf q}_{01},{\bf q}'_{01}).
 \end{eqnarray}

 All the relevant information for shot noise
 is contained in the diagonal $S_{\alpha \alpha}$, $\alpha=L,R$.
 Shot noise is obtained by calculating 
 this latter expression in the limit of low temperature but 
 finite voltage bias $V$ between the two leads.
 In that case, it is easily checked that the contribution to
 $B$ arising from the
 first term on the right-hand side of Eq.~(\ref{Btensor}) does not
 contribute, and one gets
 \begin{eqnarray}\label{shot-noise}
 S_{\alpha \alpha}(0) &=& \frac{e^2}{h} \int {\rm d}E \; \sum_{\beta \ne \gamma}
 \int 
 \rmd{\bf q}_3 \rmd{\bf q}_1 \rmd{\bf q}_{03} \rmd{\bf q}_{01} \rmd{\bf q}'_{01}
 \sum_{m,n} \sum_{m',n'}
 \Big({\mathbb S}_{\alpha \beta; m'm}({\bf q}_3,{\bf q}'_{01};E) \Big)^\dagger \;
 {\mathbb S}_{\alpha \gamma; m'n}({\bf q}_3,{\bf q}_{03};E) \nonumber \\
 && \times
 \Big({\mathbb S}_{\alpha \gamma; n'n}({\bf q}_1,{\bf q}_{03};E) \Big)^\dagger 
 \;
 {\mathbb S}_{\alpha \beta; n' m}({\bf q}_1,{\bf q}_{01};E) \;
 \Big( f_\beta(E) [1\mp f_\gamma(E)] + [1\mp f_\alpha(E)] f_\beta(E) \Big)
 \eta_{\rm env}({\bf q}_{01},{\bf q}'_{01}).\qquad
 \end{eqnarray}
 We finally assume a slow dependence of ${\mathbb S}$ on $E$, in which case
 the integral over the energy is easily performed, giving a factor $eV$. 
 For a two-lead device (L,R), we find that 
 \begin{eqnarray}
 \label{shot-noise-final}
 S_{\rm RR}(0) &=& \frac{2 e^3 V}{h} 
 \int 
 \rmd{\bf q}_3 \rmd{\bf q}_1 \rmd{\bf q}_{03} \rmd{\bf q}_{01} \rmd{\bf q}'_{01}
 \sum_{m \in L} \; \sum_{n, m',n' \in R}
 \Big({\mathbb S}_{RL; m'm}({\bf q}_3,{\bf q}'_{01};E) \Big)^\dagger \;
 {\mathbb S}_{RR; m' n}({\bf q}_3,{\bf q}_{03};E) 
 \nonumber \\
 && \qquad \qquad \qquad \times
 \Big({\mathbb S}_{RR; n'n}({\bf q}_1,{\bf q}_{03};E) \Big)^\dagger 
 \;
 {\mathbb S}_{RL; n' m}({\bf q}_1,{\bf q}_{01};E) \;
 \eta_{\rm env}({\bf q}_{01},{\bf q}'_{01}).
 \end{eqnarray}

 \end{widetext}

%~~~~~~~~~~~~~~~~~~~~~~~~~~~~thebibliography ~~~~~~~~~~~~~~~~~~~~~~~~~~~~~~~%


\begin{thebibliography}{99}
\bibitem{caveatucf} At finite $\tE$, the ergodic hypothesis breaks down and
sample-to-sample conductance fluctuations are dominated by classical effects.
They lose their universality and increase by orders of magnitude as
$\tE/\tD$ becomes sizeable. See Refs.~\cite{JacqSukho04,Two04}.
\bibitem{caveat-time} 
More precisely, the encounter introduces a typical length scale  
$\delta r_{\perp}$, that corresponds to the perpendicular distance between 
the two paths in the vicinity  of the encounter. 
Assuming hyperbolic dynamics,  we get  
$\delta r_{\perp} =\frac{v_{\rm F}\epsilon}{2\lambda}\sim L\epsilon $, 
and the exact definition of a typical cut-off  time is given by  
$T_{\ell }(\epsilon) = \lambda^{-1} \ln [({\ell} /\delta r_{\perp})^2] $, 
with  ${\ell}  =\{ {\rm L, \, W,\, }\xi \}$.
\bibitem{footnote:shot-noise-time-shift} 
This was not the case in Ref.~[\onlinecite{wj2005-fano}] 
where we used the time-independence of the problem
to shift paths $\gamma2,\gamma3$ in time, so that all paths entered
the cavity at the same time.  Such a shift is unhelpful when the
environment is present, so we do not do it here.
\bibitem{footnote:integrals}
For  $\de r_\perp \ll \lambda_{\rm F}$, we evaluate the integral
by approximating $\e^{\rmi x}\simeq 1$.
For   $\lambda_{\rm F} \ll \de r_\perp \ll L_\om$,
we split the integral into two at an arbitrary value $x'$, where 
$(\de r_\perp/L_\om) \ll x'\ll 1$.  For the integral from
$\de r_\perp/L$ to $x'$ we approximate  $\e^{\rmi x}\simeq 1$,
while for the integral from $x'$ to  $p_{\rm F}\de r_\perp/\hbar$
we approximate the denominator by $x$.
In the latter case we get an exponential integral, ${\rm Ei}(z)$,
whose properties at $z \ll 1$ and $z \gg 1$ are well-known.
The arbitrary cut-off, $x'$, drops out of the final result,
as it should.
Finally for $L_\om \ll \de r_\perp \sim  L$,
we can use integration-by-parts 
to perform an asymptotic expansion for large $\de r_\perp/L_\om$.
\end{thebibliography}
\end{document}